\documentclass[prd,aps,superscriptaddress,nofootinbib,amsmath,amssymb,9pt]{revtex4}
\usepackage{graphicx}
\usepackage{braket}
\usepackage{nicefrac}
\usepackage{float}%
\usepackage{amsmath}
\usepackage{slashed}
\usepackage{hyperref}
\usepackage{bm}
\usepackage{multirow}
 \usepackage{bbold} 

\usepackage[usenames, dvipsnames]{color}

\def\slashchar#1{\setbox0=\hbox{$#1$}
   \dimen0=\wd0 \setbox1=\hbox{/} \dimen1=\wd1
   \ifdim\dimen0>\dimen1 \rlap{\hbox to \dimen0{\hfil/\hfil}} #1
   \else  \rlap{\hbox to \dimen1{\hfil$#1$\hfil}} / \fi}

\graphicspath{{./plots/}}

\begin{document}
\title{Charged current weak production of $\Delta(1232)$ induced by electrons and positrons}

\author{A. \surname{Fatima}}
\affiliation{Department of Physics, Aligarh Muslim University, Aligarh-202 002, India}
\author{M. Sajjad \surname{Athar}}
\email{sajathar@gmail.com} 
\affiliation{Department of Physics, Aligarh Muslim University, Aligarh-202 002, India}
\author{S. K. \surname{Singh}}
\affiliation{Department of Physics, Aligarh Muslim University, Aligarh-202 002, India}

\begin{abstract} 
The charged current weak production of $\Delta (1232)$ from the free proton target induced by the electron/positron in the intermediate energy range corresponding to the beam energy available at JLab and Mainz, has been studied. 
The results for the differential scattering cross section $\frac{d\sigma}{dQ^2}$, the angular distribution $\frac{d\sigma}{d\Omega_{\Delta}}$, and the total scattering cross section $\sigma(E_e)$ for both the electron and positron induced processes are presented, for the various energies in the range of 0.5--4~GeV. 
The cross section $\sigma(E_e)$ is found  to be of the order of $10^{-39}$~cm$^{2}$ for the electron/positron energies in the few GeV range.
The availability of electron/positron beams having well defined energy and direction with very high luminosity of the order of $10^{38}-10^{39}$~cm$^{-2}$~sec$^{-1}$, makes it possible to observe the weak charged current production of $\Delta(1232)$ and determine the axial vector form factors $C_{i}^{A} (Q^2);~(i=3-5)$. 
The sensitivity of the differential cross section $\frac{d\sigma}{dQ^2}$ to the subdominant form factors $C_{3}^{A}(Q^2)$ and $C_{4}^{A} (Q^2)$ is found to be strong enough, especially in the low $Q^2$ region, which can be used to determine them phenomenologically and  to test the various theoretical models proposed to calculate them. 
\end{abstract}
\maketitle

\section{Introduction}
In the precision era of neutrino oscillation experiments, it is required to reduce the systematic errors to 1\% level~\cite{DUNE:2020ypp} so that the parameters of the PMNS mixing matrix, neutrino mass hierarchy, and CP violating phase may be determined with very high precision. 
The two major sources of systematic errors in the present experiments arise from the uncertainties in the reconstruction of the neutrino beam energy in the energy dependence of the incident neutrino flux and from the uncertainties in extracting the (anti)neutrino-nucleon scattering amplitudes from the observed data on (anti)neutrino-nucleus scattering cross sections as most of the experiments are done using moderate to heavy nuclear targets. 
In the first case, the uncertainties arise when one tries to reconstruct the energy of the incident neutrino from the measurement of the energy and angle of the final state charged lepton that has scattered quasielastically from the off-shell nucleon, which is moving with a Fermi momentum inside the nucleus. 
It has been estimated that an energy spread of around 100~MeV can occur in the reconstruction of the (anti)neutrino energy for the scattering of 1~GeV (anti)neutrino from a nucleon in the $^{12}$C nucleus~\cite{Leitner:2010kp}. 
Moreover, if one considers all other reaction channels besides the pure quasielastic scattering then an energy shift of about 100--200~MeV may occur around $E_{\nu} \sim 1$~GeV~\cite{Lalakulich:2012ac}.
In the second case, the uncertainties arise due to the various nuclear medium effects~(NME) on the cross sections when the (anti)neutrinos scatter from the bound nucleons, which are moving in a nuclear potential with a Fermi momentum and are interacting strongly with the virtual mesons and other nucleons in the nucleus giving rise to the meson exchange currents~(MEC) and other multinucleon correlation effects. 
The effects due to the binding energy and the Fermi momentum of the nucleons are relatively easy to calculate but the theoretical estimates of the nuclear medium effects due to MEC and the nucleon-nucleon correlations are found to be model dependent and contribute significantly to the systematic errors. 
It is, therefore, desirable to have the experimental data on the (anti)neutrino scattering from the hydrogen and deuterium targets~\cite{Alvarez-Ruso:2022ctb}. 
The uncertainties due to the NME are eliminated in the case of the hydrogen target and are expected to be minimal if the deuterium target is used, as the deuteron nucleus has a simple structure, which is well studied. 
But the uncertainties arising due to the reconstruction of (anti)neutrino energy in determining the energy dependence of the incident neutrino flux, where the neutrino energy band is widely spread, would still remain and lead to the systematic errors. 
Indeed, the analysis of the earliest (anti)neutrino experiments done at ANL~\cite{Barish:1978pj, Radecky:1981fn} and BNL~\cite{Kitagaki:1986ct, Kitagaki:1990vs, Baker:1980pj, Baker:1981su} using the hydrogen and deuterium targets  reflect these uncertainties in the large error bars in their results reported for the total cross section~$\sigma(E_{\nu})$ and the differential cross sections $\frac{d\sigma}{dQ^2}$.
Moreover, the data from the ANL and BNL experiments are not in agreement with each other, which is attributed predominantly due to the uncertainties in determining the energy dependence of their (anti)neutrino fluxes~\cite{Lalakulich:2005cs, Wascko:2006tx}. 
These uncertainties lead to the systematic errors in the extraction of the (anti)neutrino-nucleon scattering cross sections and the determination of the scattering amplitudes especially in the case of the quasielastic and the  inelastic reactions. 

In the energy region of around 1~GeV relevant for many of the present neutrino oscillation experiments, the inelastic reaction in which a single pion is produced, makes significant contribution to the (anti)neutrino-nucleon cross sections and is the subject of our present study. 
The single pion production~(SPP) is dominated by the process of the $\Delta(1232)$ excitation and its subsequent decays into pions and nucleons. 
The inelastic reaction of single pion production is, therefore, described by the weak $N-\Delta$ transition amplitude, which is defined in terms of the isovector vector and axial vector form factors of the $N-\Delta$ transition induced by the weak hadronic current.

The weak $N-\Delta$ transition amplitude is expressed in terms of the four isovector vector and four isovector axial vector form factors generally represented by $C_{i}^{V} (Q^2); (i=3-6)$ and $C_{i}^{A} (Q^2); (i=3-6)$, respectively, as defined in Eq.~(\ref{vec_tra_current}) of Section~\ref{formalism}~\cite{LlewellynSmith:1971uhs, Schreiner:1973mj}.
The isovector vector form factors $C_{i}^{V} (Q^2); (i=3-5)$~($C_{6}^{V} (Q^2)=0$ due to the conserved vector current~(CVC) hypothesis) are determined using the isospin symmetry and the CVC hypothesis 
for the vector current to relate them to the electromagnetic $N-\Delta$ transition form factors, which, in turn, are obtained from the experimental results on the $\gamma p \rightarrow \gamma \Delta$ and $ep \rightarrow e\Delta$ excitations. 
The isovector axial vector $N-\Delta$ transition form factors $C_{i}^{A} (Q^2); (i=3-5)$~(as $C_{6}^{A} (Q^2)$ does not contribute in the limit of $m_{l} \rightarrow 0$) are phenomenologically determined by fitting the experimental data on $\sigma(E_{\nu})$ and $\frac{d\sigma}{dQ^2}$ from the (anti)neutrino scattering from the hydrogen and deuterium targets from the ANL~\cite{Barish:1978pj, Radecky:1981fn} and BNL~\cite{Kitagaki:1986ct, Kitagaki:1990vs} experiments, making certain assumptions about $C_{i}^{A}(Q^2=0);~(i=3-5)$.
These assumptions have little theoretical justification except in the case of $C_{5}^{A}(0)$ for which the value predicted by the hypothesis of the  partial conservation of axial vector current~(PCAC) and the generalized Goldberger-Treiman~(GT) relation is used. 
In the case of $C_{3}^{A} (0)$ and $C_{4}^{A} (0)$, the values consistent with the predictions of some simple quark models based on the SU(6) symmetry and also  suggested by the Adler's model~\cite{Adler:1968tw} ($C_{3}^{A} (0)= 0$
 and $C_{4}^{A} (0)= -0.3$) using the parameterization of Schreiner and von Hippel~\cite{Schreiner:1973mj}, have been used in the literature.
For the $Q^2$ dependence of these form factors, a dipole form modified by a slowly varying function of $Q^2$ has been used to determine the axial dipole mass $M_{A}$ appearing in the definition of the dipole form factor. 
 However, the determination of the axial vector $N-\Delta$ transition form factors $C_{i}^{A} (Q^2);~(i=3-5)$ from the earlier analyses of the ANL and BNL data is not very satisfactory~\cite{Schreiner:1973mj, LlewellynSmith:1971uhs} mainly due to the uncertainties in determining the (anti)neutrino fluxes leading to the different values of the axial dipole mass $M_{A}$ from the analyses of these experiments.

In order to understand the effect of the neutrino flux uncertainties in the cross sections by the ANL and BNL groups, these experimental data have been reanalyzed by Graczyk et al.~\cite{Graczyk:2009qm, Graczyk:2009zh} and Wilkinson et al.~\cite{Wilkinson:2014yfa, Rodrigues:2016xjj}. 
Graczyk et al.~\cite{Graczyk:2009qm, Graczyk:2009zh} have taken different values for the overall normalization of the ANL and BNL neutrino fluxes, and also included the deuteron effects, which are shown to be important in the region of low $Q^2$~\cite{Alvarez-Ruso:1998ais}. 
Using different overall normalization factor of $N=1.08$ for the ANL data and $N=0.98$ for the BNL data, they find the single pion production cross section $\sigma(E_{\nu})$ to be consistent with each other within the statistical errors. 
Wilkinson et al.~\cite{Wilkinson:2014yfa, Rodrigues:2016xjj} make use of the experimental data published by the ANL and BNL collaborations on the energy dependence of the cross sections for the single pion production and the quasielastic processes i.e. $\sigma^{SPP} (E_{\nu})$ and $\sigma^{QE} (E_{\nu})$ from the neutrino-deuterium scattering to calculate the ratio $R(E_{\nu}) = \frac{\sigma^{SPP} (E_{\nu})}{\sigma^{QE} (E_{\nu})}$. 
The ratio $R(E_{\nu})$ is, therefore, free from the uncertainties arising due to the overall normalization of the neutrino fluxes. 
Indeed, the values of $R(E_{\nu})$ obtained through this procedure for the ANL and BNL experiments are found to be in agreement with each other within the statistical errors. 
The energy dependence of the cross section for the single pion production is then reproduced by multiplying this ratio in the case of each experiment by the energy dependence of the quasielastic cross section predicted by GENIE Monte Carlo event generator using $M_{A}=0.99$~GeV for the axial dipole mass in the parameterization of the nucleon axial vector form factor $g_{A}(Q^2)$. 
The GENIE prediction for the quasielastic cross section for the neutrino scattering from the deuterium target is used as it reproduces quite satisfactorily the experimental cross sections reported by the ANL and BNL collaborations for the quasielastic neutrino-deuterium scattering. 
Consequently, the energy dependence of the cross section for the single pion production from the ANL and BNL are found to be in good agreement with each other within statistical errors. 
However, to the best of our knowledge, the reanalyzed cross sections for the single pion production from these experiments have not been used to determine independently the various $N-\Delta$ transition form factors $C_{i}^{A}(Q^2); ~(i=3-5)$.

Most of the phenomenological analyses to determine the $N-\Delta$ transition form factors have focussed on the determination of $C_{5}^{A}(Q^2)$ assuming the values of $C_{3}^{A}(Q^2)$ and $C_{4}^{A}(Q^2)$ as suggested by the Adler's model~\cite{Lalakulich:2006sw, Lalakulich:2005cs, Paschos:2003qr}. 
Not much work has been done to phenomenologically determine all the three form factors $C_{3}^{A}(Q^2)$, $C_{4}^{A}(Q^2)$, and $C_{5}^{A}(Q^2)$ independently from the analysis of ANL and BNL data from the hydrogen and deuterium targets except the works of Barish et al.~\cite{Barish:1978pj}, Hemmert et al.~\cite{Hemmert:1994ky}, and Graczyk et al.~\cite{Graczyk:2009zh}. 
Barish et al.~\cite{Barish:1978pj} and Hemmert et al.~\cite{Hemmert:1994ky} use the data on $\frac{d\sigma}{dQ^2}$ in the low $Q^{2}$ region of $0<Q^2<0.1$~GeV$^{2}$ to determine the sensitivity of $\frac{d\sigma}{dQ^2}$ to different set of values of $C_{3}^{A}(0)$ and $C_{4}^{A}(0)$. 
While Barish et al.~\cite{Barish:1978pj} find the values suggested by Adler's model to be consistent with the data, however, some other set of values are also not ruled out.
The analysis of HHM model~\cite{Hemmert:1994ky} using $C_{3}^{A}(0) = 0$ find a value of $C_{4}^{A}(0) =- 0.46$, which is in disagreement with the value suggested by the Adler's model~\cite{Adler:1968tw}. 
Both the analyses do not include the deuterium effects, which were later shown to be important in the region of low $Q^2$ relevant in the analysis performed by these authors. 
In the case of Graczyk et al.~\cite{Graczyk:2009zh}, who use the value of $\frac{d\sigma}{dQ^2}$ in a larger range of $Q^2$ i.e. $0<Q^2<1$~GeV$^{2}$, using $C_{3}^{A}(0) = 0$ find a value of $C_{4}^{A}(0) = -0.67$, which is again in disagreement with the value suggested by the Adler's model~\cite{Adler:1968tw}.
Since the value of $C_{3}^{A}(0)$ is assumed to be zero in most of the analyses, therefore, there exists no experimental information on $C_{3}^{A}(Q^2)$. 
It may, therefore, be stated that the present knowledge about the form factors $C_{3}^{A}(Q^2)$ and $C_{4}^{A}(Q^2)$ obtained from the neutrino experiments is far from satisfactory and precise experimental data are needed on the hydrogen/deuterium targets for their determination. 

Theoretically, there are many calculations for the $N-\Delta$ transition form factors made in various models proposed for the nucleon structure in literature~\cite{Liu:1995bu, Mukhopadhyay:1998mn, Barquilla-Cano:2007vds, Hemmert:1994ky, Golli:2002wy, Procura:2008ze, Alexandrou:2006mc, Alexandrou:2007eyf, Chen:2023zhh, Yin:2023kom, Segovia:2014aza, Kucukarslan:2015urd, Aliev:2007pi, Unal:2021byi, Geng:2008bm}, some of which do not support the assumptions made in the Adler's model for $C_{3,4}(0)$ i.e. $C_{3}^{A}(0) = 0$ and $C_{4}^{A}(0) = - \frac{1}{4} C_{5}^{A}(0)$. 
A list of the predictions made in some of these models along with the phenomenological values determined from the experiments is given in Section~\ref{formalism}. 

In view of the difficulties associated with performing new neutrino experiments using large volume detectors filled with hydrogen and deuterium targets~\cite{Alvarez-Ruso:2022ctb} and the uncertainties associated with the wide energy spread  in the neutrino flux as discussed above, we propose to study the weak production of $\Delta(1232)$ resonance induced by the electrons and positrons through the process of single pion production, in order to determine the weak $N-\Delta$ axial vector transition form factors. 
Presently, the electron beams are available with well defined energy and direction with a good knowledge of their luminosity.
Moreover, there is possibility of the availability of the positron beams, in future, especially at JLab~\cite{Afanasev:2019xmr}, where the positron beams will have their energy, direction, and luminosity determined with same precision as the electron beams. 
Therefore, the study of the weak interaction processes induced by the electron and positron beams would eliminate the systematic error arising due to the  uncertainties in determining the incident beam fluxes as encountered in the case of (anti)neutrino experiments~\cite{Ankowski:2022thw}. 


The electron beams have traditionally been used to study the electromagnetic structure of the nucleon through the processes of elastic, inelastic and deep inelastic scattering. 
In the specific case of inelastic scattering, the electron induced excitation of 
$\Delta(1232)$ resonance has been extensively used to study the $N-\Delta$ transition form factors in the electromagnetic sector~\cite{Liu:1995bu}. 
Now, with the availability of intense electron beams with very high luminosity of the order of $10^{39}$~cm$^{-2}$~sec$^{-1}$ at JLab~\cite{Accardi:2023chb, Arrington:2021alx, JeffersonLabSoLID:2022iod, Rode:2010zz} and Mainz~\cite{Deshpande:2005wd, ELBA, Mornacchi:2023oir}, it may be feasible to observe the weak production of $\Delta(1232)$ resonance and study the $N-\Delta$ transition form factors in the CC and NC sectors of the weak interaction. 
Indeed, the weak excitation of $\Delta (1232)$ resonance induced by electrons has already been observed in the NC sector in the scattering of the  polarized electrons in the $\vec{e} p \rightarrow \vec{e} \Delta^{+}$ and $\vec{e} d \rightarrow \vec{e} \Delta^{+} n$ processes at JLab~\cite{G0:2011aa, G0:2011rpu, G0:2012hkh}, where the parity violating asymmetry arising due to the interference of the contribution due to the NC current induced by the $Z$ exchange diagram with the contribution from the parity conserving photon exchange diagram has been measured. 
The observation of the purely NC induced reactions is extremely difficult as they are highly suppressed as compared to the electromagnetic contribution as both the interactions lead to the same final states.
Moreover, there has been no experimental effort, to our knowledge, to observe the weak excitation of $\Delta (1232)$ induced by the electrons in the charged current sector.

Theoretically, there have been many calculations of the NC excitation of $\Delta(1232)$ induced by the polarized electrons and their contribution to the parity violating asymmetry~\cite{Cahn:1977uu, Mukhopadhyay:1998mn, Hammer:1995dea, Jones:1979aj, Pollock:1998tz}. 
On the other hand, there are very few calculations of the CC excitation of $\Delta(1232)$ induced by the electrons. 
In view of this, it may be worthwhile to explore the feasibility of experimentally studying the charged current production of $\Delta (1232)$ resonance induced by the electrons and positrons on the proton target through the reactions:  
\begin{eqnarray}\label{eq:delta}
 e^{-} + p &\longrightarrow& \Delta^{0} + \nu_{e}, \qquad \quad \text{ and }  \qquad \quad \\
 \label{eq:delta2}
 e^{+} + p &\longrightarrow& \Delta^{++} + \bar{\nu}_{e}.
\end{eqnarray}
These weak processes were theoretically studied almost 40 years ago by Hwang et al.~\cite{Hwang:1987sd} and later by Alvarez-Ruso et al.~\cite{AlvarezRuso:1997jr}. While Hwang et al.~\cite{Hwang:1987sd} used a constituent quark model to evaluate the $N-\Delta$ transition form factors and the cross sections, 
Alvarez-Ruso et al.~\cite{AlvarezRuso:1997jr} calculated the cross sections in a model independent way using $N-\Delta$ transition form factors determined from the 
experimental data on the neutrino scattering.
A brief summary of the status of the present information about the weak $N-\Delta$ transition form factor known from the theoretical and experimental studies of neutrino scattering is presented in Section~\ref{formalism}.

Our focus in the present work is on the weak charged current production of $\Delta(1232)$ induced by the electron/positron in order to study the axial vector form factors in the $N-\Delta$ transition. 
This is because in 
the case of the NC excitation of $\Delta(1232)$ studied through the measurement of the electron asymmetry, it has been found that the electron asymmetry is not very sensitive to the contribution from the axial vector form factors of the $N-\Delta$ transition~\cite{Mukhopadhyay:1998mn, Nath:1979qe}. 
In fact, the model calculations show that the dominant contribution to the asymmetry~($\sim 93\%$)~\cite{G0:2011aa, G0:2011rpu, G0:2012hkh} comes from the almost model independent term involving the isovector vector part of the $N-\Delta$ transition, which is multiplied by the axial vector coupling of the electron to the $Z$ boson. The contribution from the isovector axial vector part of the $N-\Delta$ transition, which is multiplied by the vector coupling of the electron to the $Z$ boson is very small i.e. only 5.2\%~\cite{G0:2011aa, G0:2011rpu, G0:2012hkh}. 
The remaining contribution comes from the nonresonant terms. 
So the electron asymmetry measurements are not sensitive enough to determine the axial vector form factor of the $N-\Delta$ transitions. 
However, the small contribution from the axial vector form factors to the parity violating electron asymmetry is consistent with the values of the axial vector form factors $C_{3}^{A} (Q^2)$, $C_{4}^{A} (Q^2)$ and $C_{5}^{A} (Q^2)$
used in the analysis of the neutrino experiments. 
It is, therefore, desirable that the pure CC weak excitation of $\Delta$ induced by the electrons and positrons is pursued to study the weak $N-\Delta$ transitions.

The theoretical calculations of the weak $N-\Delta$ transition form factors has a long history~\cite{LlewellynSmith:1971uhs}. 
These calculations have been done using various versions of the quark models of the nucleon structure~\cite{Liu:1995bu, Mukhopadhyay:1998mn, Barquilla-Cano:2007vds, Hemmert:1994ky, Golli:2002wy, Procura:2008ze}, the lattice gauge theories~\cite{Alexandrou:2006mc, Alexandrou:2007eyf, Chen:2023zhh, Yin:2023kom, Segovia:2014aza}, the QCD sum rules~\cite{Kucukarslan:2015urd, Aliev:2007pi}, and the chiral perturbation theory~\cite{Unal:2021byi, Geng:2008bm}, etc. 
Since the vector form factors in the $N-\Delta$ transitions are phenomenologically determined from the photon and electron scattering experiments, the recent studies in the various models focus mainly on the calculation of the isovector axial vector $N-\Delta$ transition form factors $C_{i}^{A}(Q^2); (i=3-6)$ and the comparison of $C_{5}^{A}(0)$ with the prediction of PCAC and the generalized GT relation. 

In view of the large systematic uncertainties in the neutrino data, it is not possible to validate the assumptions made in the case of $C_{3}^{A} (Q^2)$ and  $C_{4}^{A} (Q^{2})$ as their effect is not too large in the (anti)neutrino cross sections. 
In the absence of high precision data on neutrino scattering or the data on the electron asymmetry measurements in the electron induced NC excitation of $\Delta(1232)$, the study of CC induced weak production of $\Delta(1232)$ induced by the electrons and positrons becomes highly desirable in order to study the subdominant axial vector form factors  $C_{3}^{A} (Q^2)$ and  $C_{4}^{A} (Q^{2})$.

In this work, the weak charged current production of $\Delta (1232)$ resonance induced by electrons and positrons through the reactions given in Eqs.~(\ref{eq:delta}) and (\ref{eq:delta2}) is studied, using all the isovector vector $C_{i}^{V} (Q^2)$  and axial vector $C_{i}^{A} (Q^2)$ $N-\Delta$ transition form factors. 
In Section~\ref{formalism}, we define the $N-\Delta$ transition matrix element in terms of these form factors and discuss the present status of our knowledge about these form factors. 
In Section~\ref{results}, we give the results for the total cross section $\sigma(E_{e})$ as a function of the energy of the incoming lepton, and $\frac{d\sigma}{dQ^2}$ vs. $Q^2$ and $\frac{d\sigma}{d\Omega_{\Delta}}$ vs. $\cos\theta_{\Delta}$ for some energies relevant for JLab and discuss the feasibility of measuring them in view of the advances made in achieving highly intense electron and positron beams at JLab. 
Finally in Section~\ref{summary}, we give a summary and conclusions of this work.

\section{$N-\Delta$ transition matrix element and form factors}\label{formalism}
 \subsection{Transition matrix element}\label{sec:matrix_element}
 
The production process of the $\Delta$ resonance induced by the charged current reaction of an electron/positron from the free proton target is represented as:
  \begin{eqnarray}\label{delta}
    e^-(k) + p(p) &\longrightarrow& \nu_e(k^\prime) + \Delta^0(p^\prime), \\
    \label{delta2}
    e^+(k) + p(p) &\longrightarrow& \bar{\nu}_e(k^\prime) + \Delta^{++}(p^\prime),
  \end{eqnarray}
  where the quantities in the parentheses represent the four momenta of the corresponding particles.
  
 The transition matrix elements for the reactions given in Eqs.~(\ref{delta}) and (\ref{delta2}), respectively, are defined as
 \begin{eqnarray}\label{mat1}
  {\cal M}^{e^{-}\nu} &=& \frac{G_{F} \cos\theta_{C}}{\sqrt{2}} \; l_{\mu}^{e^{-}\nu}\: \bra{\Delta^{0} (p^{\prime})} j^{\mu} \ket{p(p)}, \\
  \label{mat2}
   {\cal M}^{e^{+}\bar{\nu}} &=& \frac{G_{F} \cos\theta_{C}}{\sqrt{2}} \; l_{\mu}^{e^{+}\bar{\nu}}\: \bra{\Delta^{++} (p^{\prime})} j^{\mu} \ket{p(p)},  
 \end{eqnarray}
where $G_{F}~(=1.166 \times 10^{-5} \text{ GeV}^{-2})$ is the Fermi coupling constant and $\theta_{C}~(=13.1^{\circ})$ is the Cabibbo mixing angle, and the leptonic current $l_{\mu}$ appearing in Eqs.~(\ref{mat1}) and (\ref{mat2}), respectively, is given by
  \begin{eqnarray}\label{lmu:e}
   l_{\mu}^{e^{-}\nu} &=& \bar{u}_{\nu}(k^{\prime}) \gamma_{\mu} \left(1-\gamma_{5} \right) u_{e^{-}} (k) \\
   \label{lmu:e+}
   l_{\mu}^{e^{+}\bar{\nu}} &=& \bar{v}_{\nu}(k^{\prime}) \gamma_{\mu} \left(1+\gamma_{5} \right) v_{e^{-}} (k) .
  \end{eqnarray}
  The matrix element for the hadronic current $j^{\mu}$ for the reaction given in Eq.~(\ref{delta}) is written as
  \begin{equation}\label{jmu}
\bra{\Delta^{0}(p^{\prime})}j^\mu \ket{p(p)}= \bar \Psi_{\beta}(p^{\prime}){ \mathcal O}^{\beta \mu} u( p).
\end{equation}
In the above expression $u(p)$ is the Dirac spinor for the proton and  $\Psi_\beta(p^{\prime})$ is a Rarita-Schwinger field for spin-$\frac{3}{2}$ particle.
 $\mathcal O^{\beta \alpha}={\mathcal O}_V^{\beta \alpha}+{\mathcal O}_A^{\beta \alpha}$ is the $N-\varDelta$ transition vertex, which is described in terms of the vector~(${\mathcal O}_V^{\beta \alpha}$)
 and the axial vector~(${\mathcal O}_A^{\beta \alpha}$) transition vertices. The vector and axial vector $N-\Delta$ transition vertices are, in turn, described in terms of the vector and axial vector form factors as~\cite{SajjadAthar:2022pjt, Athar:2020kqn}
\begin{eqnarray}\label{vec_tra_current}
{\mathcal O}_V^{\beta \alpha}&=&\left(\frac{C_{3}^V(Q^2)}{M}(g^{\alpha \beta}\not\! q-q^{\beta}\gamma^{\alpha})
+\frac{C_{4}^V(Q^2)}{M^2}(g^{\alpha \beta} q \cdot p^{\prime}-q^{\beta}p^{\prime\alpha}) + 
\frac{C_{5}^V(Q^2)}{M^2}(g^{\alpha \beta}q \cdot p-q^{\beta}p^{\alpha}) \right. \nonumber \\
&& + \left.\frac{C_{6}^{V}(Q^2)}{M^2}q^{\beta}q^{\alpha}\right)\gamma_{5} \\
\label{ax_tra_current}
{\mathcal O}_A^{\beta \alpha}&=&\left(\frac{C_{3}^A(Q^2)}{M}(g^{\alpha \beta}\not\! q-q^{\beta}\gamma^{\alpha})
+
\frac{C_{4}^{A}(Q^2)}{M^2}(g^{\alpha \beta} q \cdot p^{\prime}-q^{\beta}p^{\prime\alpha})
+C_{5}^{A}(Q^2)g^{\alpha \beta}+\frac{C_{6}^{A}(Q^2)}{M^2}q^{\beta}q^{\alpha}\right).~~~
\end{eqnarray}
In Eqs.~(\ref{vec_tra_current}) and (\ref{ax_tra_current}), $C_{i}^{V}(Q^2); (i=3-6)$ are the isovector vector and $C_{i}^{A}(Q^2); (i=3-6)$ are the isovector axial vector $N-\Delta$ transition form factors.

The vector and axial vector $N-\Delta$ transition form factors have been determined phenomenologically by analyzing the experimental data on the (anti)neutrino scattering from the hydrogen and deuterium targets by many authors~\cite{Paschos:2003qr, Lalakulich:2005cs, Hernandez:2010bx, Hernandez:2007qq, Graczyk:2009qm, Alvarez-Ruso:2015eva, Lalakulich:2006sw}. 
In the following, we describe the properties of these form factors and review, in brief, the various parameterizations of the vector and axial vector $N-\Delta$ transition form factors, which have been used in the literature to study the neutrino induced production of $\Delta(1232)$ resonance.

\subsection{$N-\Delta$ transition form factors}\label{FF}
The vector and axial vector $N-\Delta$ transition form factors are determined using some general symmetry properties of the weak hadronic current, such as:
\begin{itemize}
\item [(i)] The invariance under time reversal, which implies that all the vector~($C_{i}^{V}(Q^2); i=3-6$) and axial vector~($C_{i}^{A}(Q^2); i=3-6$) form factors must be real.

\item [(ii)] The isospin symmetry, which relates the hadronic matrix elements of $p\rightarrow \Delta^{0}$ and $p\rightarrow \Delta^{++}$ transitions via the following relation
 \begin{equation}\label{iso:delta}
 \bra{\Delta^{++}(p^{\prime})}j^\mu \ket{p(p)} = \sqrt{3} ~\bra{\Delta^{0}(p^{\prime})}j^\mu \ket{p(p)} .
 \end{equation}
\end{itemize}

\subsubsection{Vector form factors}
The isovector vector form factors are determined using the following properties of the weak vector currents: 
\begin{itemize}
 \item [(i)] The principle of conserved vector current~(CVC) hypothesis of the weak vector currents, which implies that $C_{6}^{V} (Q^2) = 0$.
 
 \item [(ii)] Isospin symmetry, which relates the weak vector form factors~($C_{i}^{V} (Q^2)$) to the electromagnetic~($C_{i}^{N} (Q^2)$) $N-\Delta$ transition form factors via the relation 
 \begin{equation}
  C_{i}^{V} (Q^2) = - C_{i}^{N} (Q^2); \qquad \quad i=3,4,5.
 \end{equation}
\end{itemize}

The earlier data on the weak production processes of $\Delta(1232)$ were analyzed assuming magnetic dipole~($M1$) dominance of the electromagnetic $p \rightarrow \Delta^{+}$ excitation, which implies~\cite{LlewellynSmith:1971uhs}
\begin{equation}\label{C3C4V}
 C_{4}^{V} (Q^2) = - \frac{M}{W} C_{3}^{V} (Q^{2}); \qquad \quad  C_{5}^{V} (Q^2)=0.
\end{equation}
However, in the light of the nonzero value of electric quadrapole to magnetic dipole~($E2/M1$) amplitudes observed in the electromagnetic excitation of $\Delta$, the vector transition form factors are now  determined from the experimentally observed helicity amplitudes $A_{\frac{1}{2}}$, $A_{\frac{3}{2}}$ and $S_{\frac{1}{2}}$ in the $ep\rightarrow e\Delta^{+}$ transition, for example, from the MAID analysis~\cite{Tiator:2011pw}.

In the original analysis of ANL~\cite{Barish:1978pj, Radecky:1981fn} and BNL~\cite{Kitagaki:1986ct, Kitagaki:1990vs} data, the following parameterizations for $C_{i}^{V}(Q^2); (i=3-5)$ were used~\cite{Dufner:1967yj}:
\begin{eqnarray}
 \left| C_{3}^{V}(Q^2) \right|^{2} &=& (2.05)^{2}\; \left[1 + 9 \sqrt{Q^2} \right]\; 
 \exp \left[-6.3 \sqrt{Q^2} \right],
\end{eqnarray}
along with Eq.~(\ref{C3C4V}) for $C_{4}^{V}(Q^2)$ and $C_{5}^{V}(Q^2)$.

However, more recently, the parameterization given by Lalakulich {et al.}~\cite{Lalakulich:2006sw} have been used in many calculations of the weak $N-\Delta$ excitations~\cite{Hernandez:2007qq, Hernandez:2010bx, Graczyk:2009qm, Graczyk:2009zh, Alvarez-Ruso:2015eva, Kakorin:2021mqo}. These are given by:
\begin{equation}\label{civ_lala}
C_i^V(Q^2)=\frac{C_i^V(0)}{\left(1+\frac{Q^2}{M_V^2}\right)^{2}}~{\cal{D}}_i,~~~i=3,4,5,
\end{equation}
with $C_3^V(0)=2.13$, $C_4^V(0)=-1.51$ and $C_5^V(0)=0.48$, and
\begin{eqnarray}\label{di} 
{\cal{D}}_{3,4}&=&\left(1+\frac{Q^2}{4M_V^2}\right)^{-1}~~~ \mbox{and}\nonumber\\
{\cal{D}}_{5}&=&\left(1+\frac{Q^2}{0.776M_V^2}\right)^{-1};~~ M_V=0.84~\rm{GeV}
\end{eqnarray} 

Theoretically, there exist many calculations of the $N-\Delta$ vector transition form factors in various models proposed for the structure of the nucleon and $\Delta$~\cite{Nath:1979qe, Mukhopadhyay:1998mn, Liu:1995bu, Hemmert:1994ky, Hernandez:2010bx, Buchmann:2001gj}. 
But the recent analyses of the neutrino scattering data to determine the weak $N-\Delta$ transitions have preferred to use the phenomenological values of these form factors obtained from the analysis of the experimental data on $\gamma p \rightarrow \gamma \Delta$ and $ep \rightarrow e\Delta$ transitions given in Eq.~(\ref{civ_lala}).

 \subsubsection{Axial vector form factors}
  The isovector axial vector form factors are determined using the following properties of the weak hadronic current.
 \begin{itemize}
  \item [(a)] {\bf $\bm{C_{5}^{A} (Q^2)}$ and $\bm{C_{6}^{A} (Q^2)}$}   

 \begin{itemize}
 \item [(i)] In the axial vector sector, the dominant contribution to the cross section comes from $C_{5}^{A}(Q^2)$, which is determined using the hypothesis of PCAC along with the pion pole dominance of the divergence
of the axial-vector current~(PDDAC), which relates the axial vector coupling $C_{5}^{A}(0)$ to the strong $\Delta \rightarrow N\pi$ coupling constant $g_{\Delta N\pi}$ through the generalized
GT relation, i.e.,
 \begin{eqnarray}\label{C5A0}
  C_5^A(0) &=& f_\pi \frac{ g_{\Delta N \pi}  }{2 \sqrt3 M },
 \end{eqnarray}
  using $f_{\pi} = 0.97m_{\pi}$ and $g_{\Delta N\pi}=28.6$, we find $C_{5}^{A} (0) = 1.2$.
  
   In Table~\ref{tab:axial_ff}, we list the various values of $C_{5}^{A}(0)$ obtained in some of the experimental and theoretical analyses taken from the existing literature on the subject. 
It may be observed from the table that there is a large variation in the theoretical prediction of $C_{5}^{A}(0)$. 
  
 \item [(ii)] Using PCAC and the generalized Goldberger-Treiman relation, the form factor $C_{6}^{A} (Q^2)$ is given in terms of $C_{5}^{A}(Q^2)$, i.e.,  
  \begin{eqnarray}\label{C6A}
  C_{6}^{A}(Q^2) &=& C_{5}^{A}(Q^2) \frac{M^{2}}{Q^2 + m_{\pi}^{2}},
 \end{eqnarray}
 with $m_{\pi}$ being the mass of the pion.
 
 The contribution of $C_{6}^{A} (Q^2)$ to the matrix element is proportional to the lepton mass and is negligible in the case of electron/positron induced reactions.

\item [(iii)] The $Q^2$ dependence of the axial vector form factor $C_{5}^{A} (Q^2)$ is given by Schreiner and von Hippel~\cite{Schreiner:1973mj} parameterization of the Adler's model~\cite{Adler:1968tw}, i.e.,
 \begin{eqnarray}\label{c5aq}
  C_{5}^A(Q^2) &=&  \frac{C_{5}^A(0) \left( 1+ \frac{a\, Q^2}{b~+~Q^2} \right)}{\left( 1 + Q^2 /M_{A}^2 \right)^2} ,
\end{eqnarray}
with $C_{5}^{A} (0)=1.2$, $a=-1.21$, and $b=2$~GeV$^{2}$ have been used in the earlier analyses.  
In the case of $C_{6}^{A}(Q^2)$, the $Q^2$ dependence is determined by Eq.~(\ref{C6A}).
 
 In recent years, many authors~\cite{Hernandez:2007qq, Hernandez:2010bx, Graczyk:2009qm, Graczyk:2009zh, Alvarez-Ruso:2015eva, Kakorin:2021mqo} have used the modified dipole parameterization for the $Q^2$ dependence of $C_5^A(Q^2)$ with different values of the axial dipole mass $M_{A}$. In the present work, we have used the parameterization of $C_5^A(Q^2)$ given by the model of Lalakulich et al.~\cite{Lalakulich:2006sw}, i.e.
\begin{eqnarray}\label{cia_lala}
C_5^A(Q^2)&=&\frac{C_5^A(0)}{\left(1+\frac{Q^2}{3M_A^2}\right)\left(1+\frac{Q^2}{M_A^2}\right)^{2}},
\end{eqnarray}
 with $C_5^A(0)=1.2$ and $M_{A}=1$~GeV.
 \end{itemize}

\begin{table*}
\centering
\begin{tabular*}{160mm}{@{\extracolsep{\fill}} c  c c c c }\hline\hline
 & Model & $C_{3}^{A}(0)$ & $C_{4}^{A}(0)$ & $C_{5}^{A}(0)$ \\ \hline
  &Adler~\cite{Adler:1968tw, Schreiner:1973mj} & 0 & $-0.3$ & 1.2 \\ 
Phenomenological &HHM~\cite{Hemmert:1994ky}  & 0 & $-0.46 \pm 0.06$ & $1.39 \pm 0.14 $ \\ 
&Graczyk et al.~\cite{Graczyk:2009zh}  & 0 & $-0.67 \pm 0.42$ & $1.17 \pm 0.13 $ \\ \hline
&Isgur-Karl~\cite{Liu:1995bu}  & 0.0008 & $-0.657$ & 1.2 \\ 
&HHM~\cite{Hemmert:1994ky}  & 0 & $-0.29 \pm 0.006$ & $0.87 \pm 0.03 $ \\ 
&SU(6)~\cite{Liu:1995bu}  & 0 & $-0.38$ & $1.17$ \\ 
Theoretical&D-mixing~\cite{Liu:1995bu}  & 0.052 & 0.052 & 0.813\\ 
&Barquilla-Cano et al.~\cite{Barquilla-Cano:2007vds} & 0.035 & $-0.26$ & 0.93 \\ 
&Golli~\cite{Golli:2002wy}  & 0 & 0.141 & 1.53 \\ 
&Kucukarslan~\cite{Kucukarslan:2015urd} (conventional) & $0.11 \pm 0.03$ & $0.27 \pm 0.09$&  $1.14 \pm 0.20$ \\ 
&Chen et al.~\cite{Chen:2023zhh}  & $0.26^{+0.17}_{-0.04}$ & $-0.66^{+0.03}_{-0.10}$  &  $1.16^{+0.09}_{-0.03}$
\\
\hline\hline
\end{tabular*}
\caption{Values of the axial vector form factors $C_{i}^{A};~(i=3-5)$ at $Q^2=0$ calculated in various phenomenological and theoretical models.}
\label{tab:axial_ff}                                                  
\end{table*}

\item [(b)] {\bf $\bm{C_{3}^{A} (Q^2)}$ and $\bm{C_{4}^{A} (Q^2)}$} \\

The subdominant axial vector form factors $C_3^A(Q^2)$ and $C_4^A(Q^2)$ are phenomenologically parameterized by Schreiner and von Hippel~\cite{Schreiner:1973mj} using the Adler's model~\cite{Adler:1968tw} as:
\begin{equation}\label{c4-c3}
      C_3^A(Q^2) =0;  \qquad     \qquad  C_4^A(Q^2) = -\frac{1}{4}C_5^A(Q^2).
\end{equation}
Lalakulich et  al.~\cite{Lalakulich:2006sw} have also used the above relation for $C_3^A(Q^2)$ and $C_4^A(Q^2)$ form factors with the modified dipole form for $C_{5}^{A}(Q^2)$ as given in Eq.~(\ref{cia_lala}), instead of Eq.~(\ref{c5aq}). 

Graczyk et al.~\cite{Graczyk:2009zh} have phenomenologically determined the form factors $C_{4}^{A}(Q^2)$ and $C_{5}^{A}(Q^2)$ independent of the Adler's model~\cite{Adler:1968tw} by assuming them to be of the dipole form, keeping  $C_{3}^{A}(Q^2)=0$, and using
\begin{equation}\label{C4;Graczyk}
 C_{i}^{A} (Q^2) = \frac{C_{i}^{A}(0)}{\left(1+\frac{Q^2}{M_{Ai}^{2}}\right)^{2}}; \qquad \quad i=4,5,
\end{equation}
and find the best fitted values of $C_{4,5}^{A}(0)$ and $M_{A4,A5}$ to be:
\begin{eqnarray}
 -C_{4}^{A}(0) &=& 0.67 \pm 0.42; \qquad \quad M_{A4} = 0.4^{+1.1}_{-0.4}~\text{GeV}, \\
  C_{5}^{A}(0) &=& 1.17 \pm 0.13; \qquad \quad M_{A5} = 0.95 \pm 0.07~\text{GeV}.
\end{eqnarray}

Theoretically, considerable attention has been given to the discussion of these axial vector $N-\Delta$ transition form factors in various versions of the quark models~\cite{Liu:1995bu, Hemmert:1994ky, Mukhopadhyay:1998mn, Procura:2008ze, Barquilla-Cano:2007vds}, the lattice QCD model~\cite{Alexandrou:2006mc, Alexandrou:2007eyf}, the chiral quark model~\cite{Golli:2002wy}, the light cone QCD sum rules~\cite{Kucukarslan:2015urd, Aliev:2007pi}, the relativistic baryon chiral perturbation theory~\cite{Geng:2008bm, Unal:2021byi}, and the QCD inspired models of the nucleon structure~\cite{Segovia:2014aza, Yin:2023kom, Chen:2023zhh}. 
It should be noted that the nonrelativistic quark model assuming exact SU(6) symmetry, predicts $C_{3}^{A} (Q^2)$ to be exactly zero and $C_{4}^{A} (Q^2)$ to be nonzero~\cite{Liu:1995bu}. 
In Table~\ref{tab:axial_ff}, we show the nonzero values of $C_{3}^{A} (0)$ and different values of $C_{4}^{A} (0)$ in the various theoretical model calculations, which reflect the different mechanisms to include minimal SU(6) symmetry breaking effects in these models. 

The models based on the light cone QCD sum rules~\cite{Kucukarslan:2015urd, Aliev:2007pi} have also reported the values for the subdominant $C_3^A(Q^2)$ and $C_4^A(Q^2)$, but this method is reliable in the region of $Q^2>1$~GeV$^{2}$. 
Since we are interested in the study of weak electroproduction of $\Delta$ at moderate electron energies in the region of $E_{e}=1-2$~GeV, where the cross sections are significant only in the low $Q^2$ region of $Q^{2} \le 1$~GeV$^{2}$, therefore, we have not used the parameterization of $C_3^A(Q^2)$ and $C_4^A(Q^2)$ obtained in the light cone QCD sum rules model, in the numerical calculations.

The subdominant axial vector form factors calculated in the chiral constituent quark model  of Barquilla-Cano et al.~\cite{Barquilla-Cano:2007vds} and the lattice gauge model of Chen et al.~\cite{Chen:2023zhh} have been used in the present work as representative for the quark model and lattice gauge theory models to study the reactions given in Eqs.~(\ref{delta}) and (\ref{delta2}).
The $Q^2$ dependence of the form factors $C_{3,4}^{A}(Q^2)$ in these models has been given numerically in their work in the form of $C_{i}^{A}(Q^2)$ vs. $Q^2$ plots, we have fitted them assuming the dipole and modified dipole forms as described below, to be used in the present work. 

The $Q^2$ dependence of $C_{4}^{A}(Q^2)$ in the model of  Barquilla-Cano et al.~\cite{Barquilla-Cano:2007vds} show similar behavior as obtained in the the Adler's model~\cite{Adler:1968tw}~(Eq.~(\ref{c4-c3})) with $C_{4}^{A}(0)=-0.26$ and $C_{5}^{A}(Q^2)$ as given in Eq.~(\ref{cia_lala}). Therefore, we parameterize the $Q^2$ dependence of $C_{3}^{A}(Q^2)$  as
\begin{equation}\label{c3:Buch}
 C_{3}^{A} (Q^2) = C_{3}^{A}(0) \frac{\left(\frac{a_{3}Q^2}{b_{3}+Q^2}\right)}{\left(1+\frac{Q^2}{M_{A3}^{2}}\right)^2}
\end{equation}
with $C_{3}^{A}(0)=0.035$, $a_{3}=-4.61$, $b_{3}=2.8$~GeV$^{2}$, and $M_{A3}=1.67$~GeV.

The $Q^2$ dependence of $C_{3}^{A}(Q^2)$ and $C_{4}^{A}(Q^2)$ as given by Chen et al.~\cite{Chen:2023zhh} is parameterized  using the dipole form as
\begin{equation}\label{c3A_Chen}
 C_{i}^{A} (Q^2) = \frac{C_{i}^{A}(0)}{\left(1+\frac{Q^2}{x_{i}M_{AS}^{2}}\right)^{2}}; \qquad \quad i=3,4
\end{equation}
with $C_{3}^{A}(0) = 0.26^{+0.17}_{-0.04}$, $C_{4}^{A}(0) = -0.66^{+0.03}_{-0.10}$, $x_{3}=1$, $x_{4}=0.74$, and $M_{AS}=1$~GeV. 
\end{itemize}

\section{Results for cross sections and discussions}\label{results}
  The general expression for the differential cross section for the processes given in Eqs.~(\ref{delta}) and (\ref{delta2}) is written as~\cite{AlvarezRuso:1997jr}
\begin{eqnarray} \label{delta_cross_section}
\frac {d \sigma}{d Q^2} = \frac{1}{128\pi^{2} M M_{\Delta}E_{e}^{2}} \int {d E_{\nu}} 
\frac{\Gamma_\Delta(W)}{(W-M_\Delta)^2+\frac{\Gamma_\Delta^2(W)}{4}} \overline{\sum} \sum |{\cal{M}}|^2,
\end{eqnarray}
where $W=\sqrt{(p+q)^2}$ is the center of mass energy of the hadronic system, with $Q^2 = -q^2$ and $q=k-k^{\prime}$ being the four momentum transfer.
$M$ and $M_{\Delta}$ are the masses of the proton and $\Delta$, respectively;  $E_{e}$ and $E_{\nu}$ are the energies of the incoming and outgoing leptons, respectively. $\Gamma_{\Delta}(W)$ is the $W$ dependent decay width of the $\Delta$ resonance.

In the expression of the differential scattering cross section given in Eq.~(\ref{delta_cross_section}), $ \overline{\sum} \sum |{\cal{M}}|^2$ is the transition matrix element squared given in terms of the leptonic and hadronic tensors as
\begin{equation}\label{mat_square}
 \overline{\sum} \sum |{\cal{M}}|^2 = \frac{G_{F}^{2} \cos^{2}\theta_{C}}{2} L_{\mu\nu} J^{\mu\nu}.
\end{equation}
The leptonic tensor $\cal{L}_{\mu\nu}$ is written in terms of the leptonic currents defined in Eqs.~(\ref{lmu:e}) and (\ref{lmu:e+}), and is given by
\begin{equation}\label{Lmunu}
 {\cal{L}_{\mu\nu}} =  \overline{\sum}\sum l_{\mu} {l_{\nu}}^{\dagger}=  4\left(k_{\mu}k_{\nu}^{\prime} - (k\cdot k^{\prime})g_{\mu\nu} + k_{\nu}k_{\mu}^{\prime} \pm i \epsilon_{\mu\nu\rho\sigma} k^{\rho}{k^{\prime}}^\sigma \right)
\end{equation}
where $+~(-)$ stands for electron~(positron) induced reactions.
The hadronic tensor 
 $J^{\mu\nu}$ is defined in terms of the hadronic current $j^{\mu}$ as 
 \begin{equation}
 J^{\mu\nu}=  \overline{\sum}\sum j^{\mu} {j^{\nu}}^{\dagger}=  \frac{1}{2}
 Tr\left[ (\slashed{p} + M) {\tilde{\mathcal O}}^{\alpha\mu } {\it P}_{\alpha \beta}
      {\mathcal O}^{\beta\nu} \right],
   \end{equation}  
   where 
   ${\it P}_{\alpha \beta}$ is the spin-3/2 projection operator given by
\begin{eqnarray}\label{propagator}
{\it P}_{\alpha \beta} = -(\not\! p^ \prime +M_{\Delta}) 
\left(g_{\alpha \beta}-\frac{2}{3} \frac{{p^\prime}_{\alpha}{p^\prime}_{\beta}}{M_{\Delta}^2}
+ \frac{1}{3}\frac{{p^\prime}_{\alpha} \gamma_{\beta}-
{p^\prime}_{\beta} \gamma_{\alpha}}{M_{\Delta}}-\frac{1}{3}\gamma_{\alpha}\gamma_{\beta}\right).
\end{eqnarray}
The hadronic current $j^{\mu}$ for $N-\Delta$ transition is given in Eq.~(\ref{jmu}),  parameterized in terms of the vector and axial vector $N-\Delta$ transition form factors, which have already been discussed in Section~\ref{FF}.

The delta decay width $\Gamma_{\Delta}(W)$ is taken as the energy dependent $P$-wave decay width given by
\begin{equation}
\Gamma_\Delta(W)=\frac{1}{6 \pi}\left(\frac{f_{\pi N \Delta}}{m_{\pi}}\right)^2 \frac{M_{\Delta}}{W}|\bm q_{cm}|^3,
\end{equation}
 where $f_{\pi N \Delta}=2.127$ is the $\Delta \rightarrow N\pi$ coupling constant, $m_{\pi}$ is the pion mass, $|\bm q_{cm}|$ is the pion momentum in the rest 
 frame of the resonance and is given by
\[|{\bm q_{cm}}|=\frac{\sqrt{(W^2-m_{\pi}^2-M^2)^2 -4 m_{\pi}^2M^2}}{2W},\] 
 where we have taken $W$ in the range $(M+m_\pi)\le W < 1.4~\rm{GeV}$.

 \subsection{Total scattering cross section~($\sigma(E_e)$) for electron and positron induced $\Delta$ production processes}
 \begin{figure}
\begin{center}
\includegraphics[width=8cm,height=8cm]{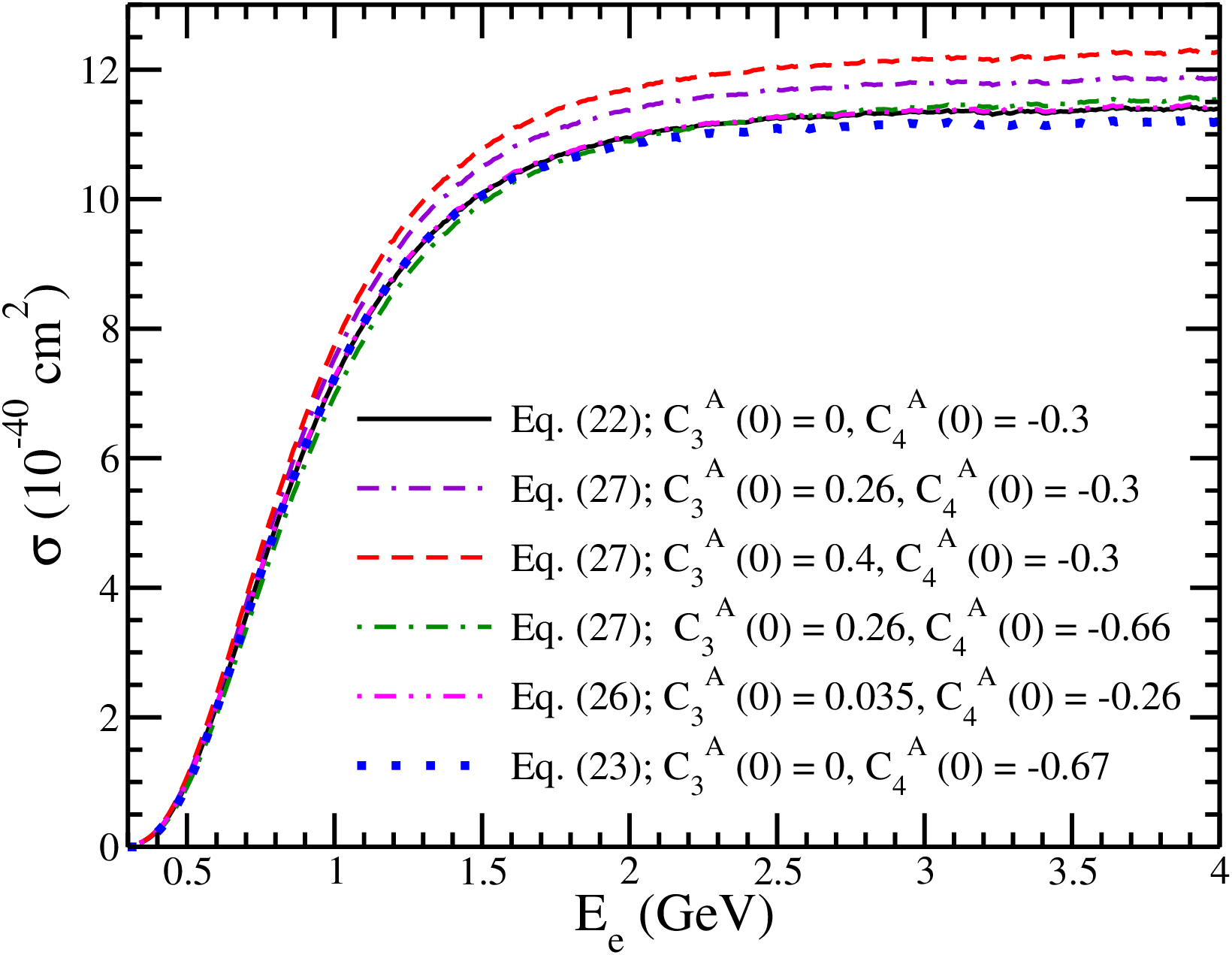}
\includegraphics[width=8cm,height=8cm]{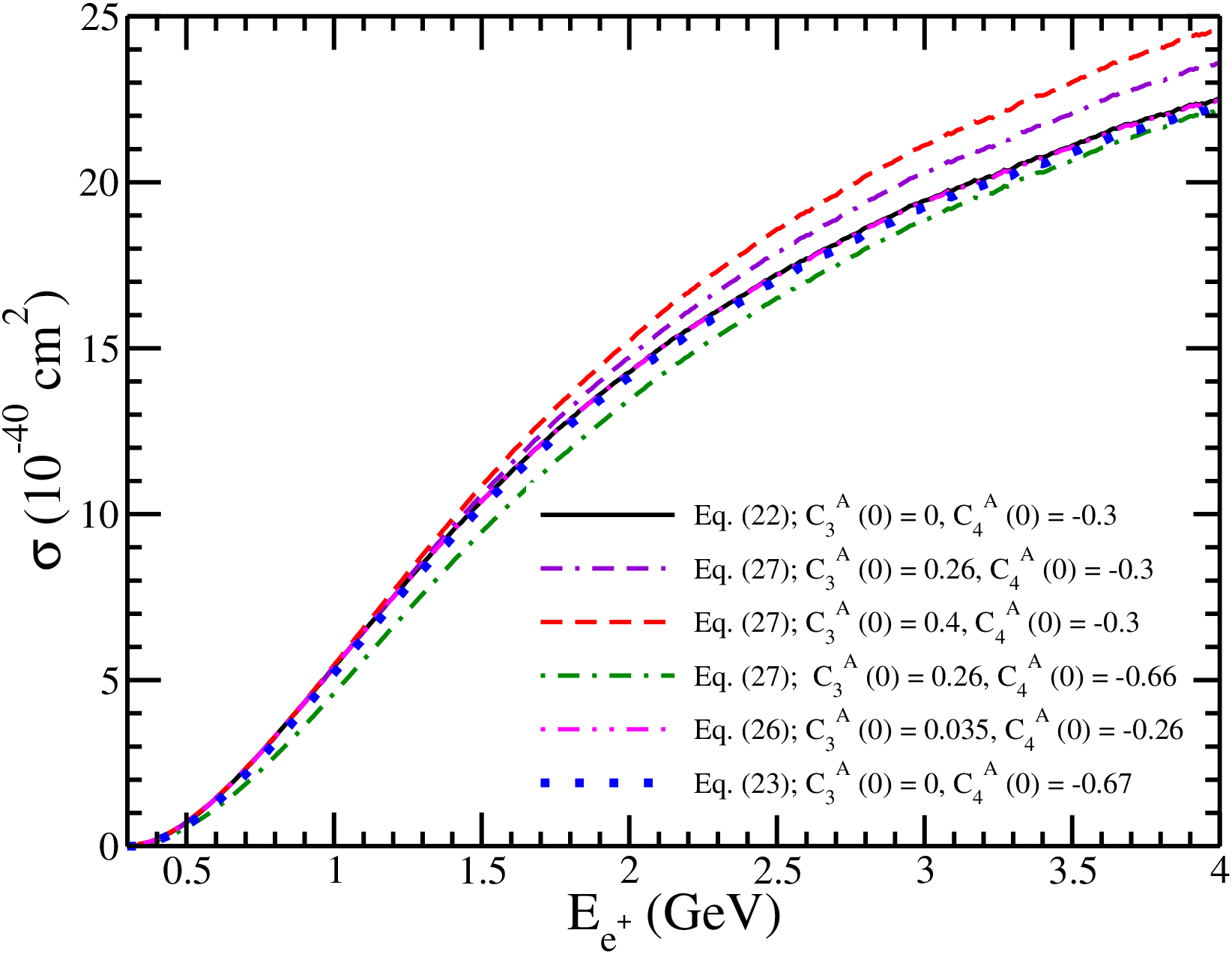}
\caption{Total scattering cross section ($\sigma$) vs electron/positron energy ($E_{e^-}/E_{e^+}$) for
 $e^- + p \longrightarrow \nu_e + \Delta^0$~(left panel) and $e^+ + p \longrightarrow \bar{\nu}_e + \Delta^{++}$~(right panel) scattering processes. 
 The solid and double-dotted-dashed lines, respectively, represent the results obtained using the form factor parameterization of Lalakulich et al.~\cite{Lalakulich:2006sw} and the modification of $C_3^{A}(Q^2)$ by Barquilla-Cano et al.~\cite{Barquilla-Cano:2007vds} (Eq.~(\ref{c3:Buch})). 
 The double-dashed-dotted line and dashed line represent the results obtained using the parameterization of $C_{3}^{A}(Q^2)$ given by Chen et al.~\cite{Chen:2023zhh} (Eq.~(\ref{c3A_Chen})) using $C_{3}^{A}(0)=0.26$ and 0.4, respectively, and for all the other vector and axial vector form factors, the parameterization given by Lalakulich et al.~\cite{Lalakulich:2006sw} is used. The dashed-dotted line represents the results obtained using the parameterization given in Eq.~(\ref{c3A_Chen}) for $C_{3}^{A}(Q^2)$ and $C_{4}^{A}(Q^2)$ for Chen et al.~\cite{Chen:2023zhh}.  
 The dotted line represents the results obtained using the parameterization given by Graczyk~\cite{Graczyk:2009zh}~(Eq.(\ref{C4;Graczyk})) for $C_{4}^{A}(Q^2)$. 
 }\label{sigma:delta_weak}
\end{center}
\end{figure}

\begin{figure}
\begin{center}
\includegraphics[width=0.47\textwidth,height=.5\textwidth]{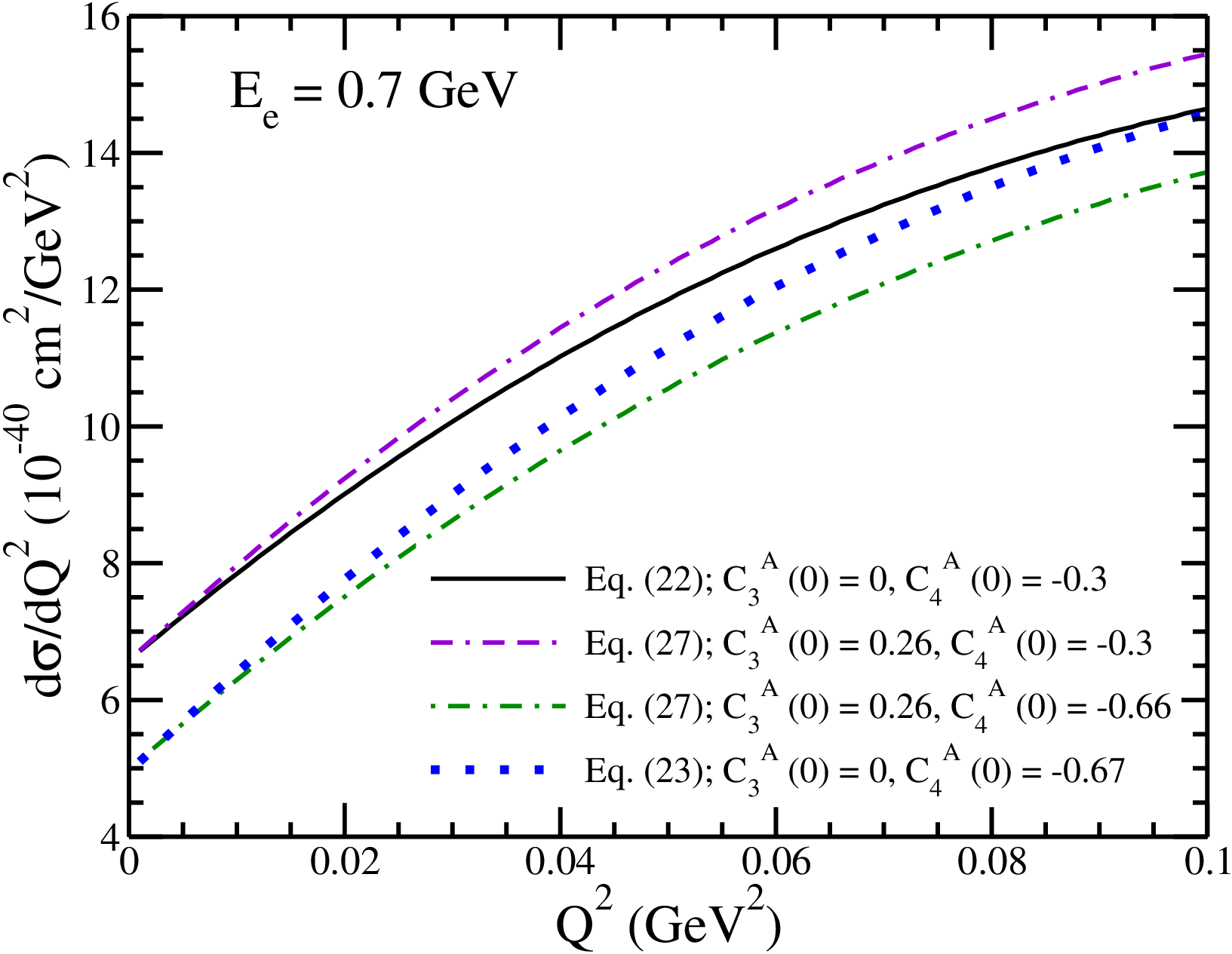}
\includegraphics[width=0.47\textwidth,height=.5\textwidth]{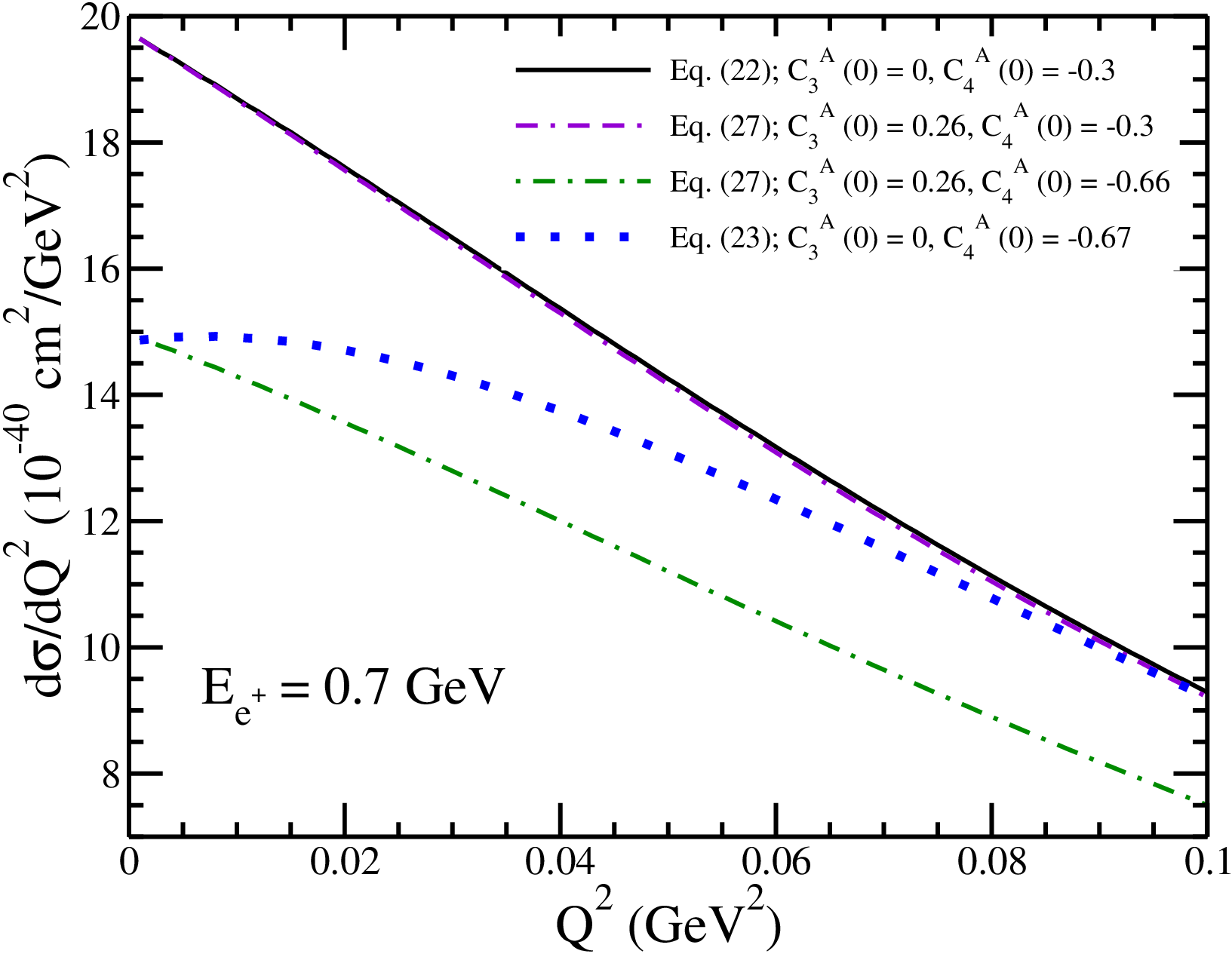}
\caption{$\frac{d\sigma}{dQ^2}$ as a function of $Q^2$ for the processes $e^- + p \longrightarrow {\nu}_e + \Delta^{0}$~(left panel) and $e^+ + p \longrightarrow \bar{\nu}_e + \Delta^{++}$~(right panel) at $E_{e}=0.7$~GeV. Lines and points have the same meaning as in Fig.~\ref{sigma:delta_weak}. }
\label{fg:dsigma_compare}
\end{center}
\end{figure}

 The total scattering cross section~($\sigma$)  for the processes $e^{-} p \rightarrow \nu_{e} \Delta^{0}$ and $e^{+}p \rightarrow \bar{\nu}_{e}\Delta^{++}$ is obtained by integrating the differential scattering cross section $\frac{d\sigma}{dQ^2}$ given in Eq.~(\ref{delta_cross_section}) over $Q^2$, using the form factors   
 $C_{i}^{V}(Q^2); (i=3-5)$ and $C_{5}^{A}(Q^2)$, given in Eqs.~(\ref{civ_lala}) and (\ref{cia_lala}), respectively. 
To study the effect of the subdominant axial vector form factors $C_{3}^{A} (Q^2)$ and $C_{4}^{A} (Q^2)$, we have used the phenomenological parameterizations given by Adler~\cite{Adler:1968tw}~(Eq.~(\ref{c4-c3})) and Graczyk et al.~\cite{Graczyk:2009zh}~(Eq.~(\ref{C4;Graczyk})), and also the parameterization obtained by us~(Eqs.~(\ref{c3:Buch}) and (\ref{c3A_Chen})) for the $Q^2$ dependence of $C_{3}^{A} (Q^2)$ and $C_{4}^{A} (Q^2)$ for the numerical values of Barquilla-Cano et al.~\cite{Barquilla-Cano:2007vds} and Chen et al.~\cite{Chen:2023zhh}.

In Fig.~\ref{sigma:delta_weak}, we present the results for the total scattering cross section $\sigma(E_{e^{-}}/E_{e^{+}})$ vs. electron/positron energy~($E_{e^-}/E_{e^{+}}$) for the electron and positron induced $\Delta$ production processes viz. $e^- + p 
\longrightarrow \nu_e + \Delta^0$ and $e^+ + p \longrightarrow \bar{\nu}_e + \Delta^{++}$. 
In the case of electron induced $\Delta$ production, the cross section increases with the increase in electron energy and saturates at around 2~GeV, while in the case of positron induced reaction, the cross section increases with the increase in positron energy even at $E_{{e}^{+}}=4$~GeV. 
The total cross section for the $\Delta^{++}$~(also in the case of $\Delta^{-}$ production) production reaction should, in principle, be larger than the cross section for the $\Delta^{+}$~(also in the case of $\Delta^{0}$ production) production process by a factor of 3 due to the isospin symmetry~(see Eq.~(\ref{iso:delta})).
However, the production cross section for the positron induced reaction, in general, are smaller than production cross section for the electron induced reaction due to the interference term of the vector and axial vector contributions. 
The combination of these two effects leads to some interesting observations on  the results of the total cross sections for the electron and positron induced channels.
In the energy region from threshold up to $E_{e}=1.5$~GeV, the production cross section for $\Delta^{0}$ induced by the electron is larger than the production cross section for $\Delta^{++}$ induced by the positron as the suppression due to the opposite sign of the interference terms is large enough to surpass the overall increase due to the isospin factor in the $\Delta^{++}$ cross section. 
As the energy of the incoming lepton increases, the contribution of the  interference terms decreases and becomes very small, therefore, the cross section for $e^+ + p \longrightarrow \bar{\nu}_e + \Delta^{++}$ process becomes larger than the production cross section for $e^- + p 
\longrightarrow \nu_e + \Delta^0$ process for energies higher than $E_{e}=1.5$~GeV.

\begin{figure}
\begin{center}
\includegraphics[width=0.47\textwidth,height=.45\textwidth]{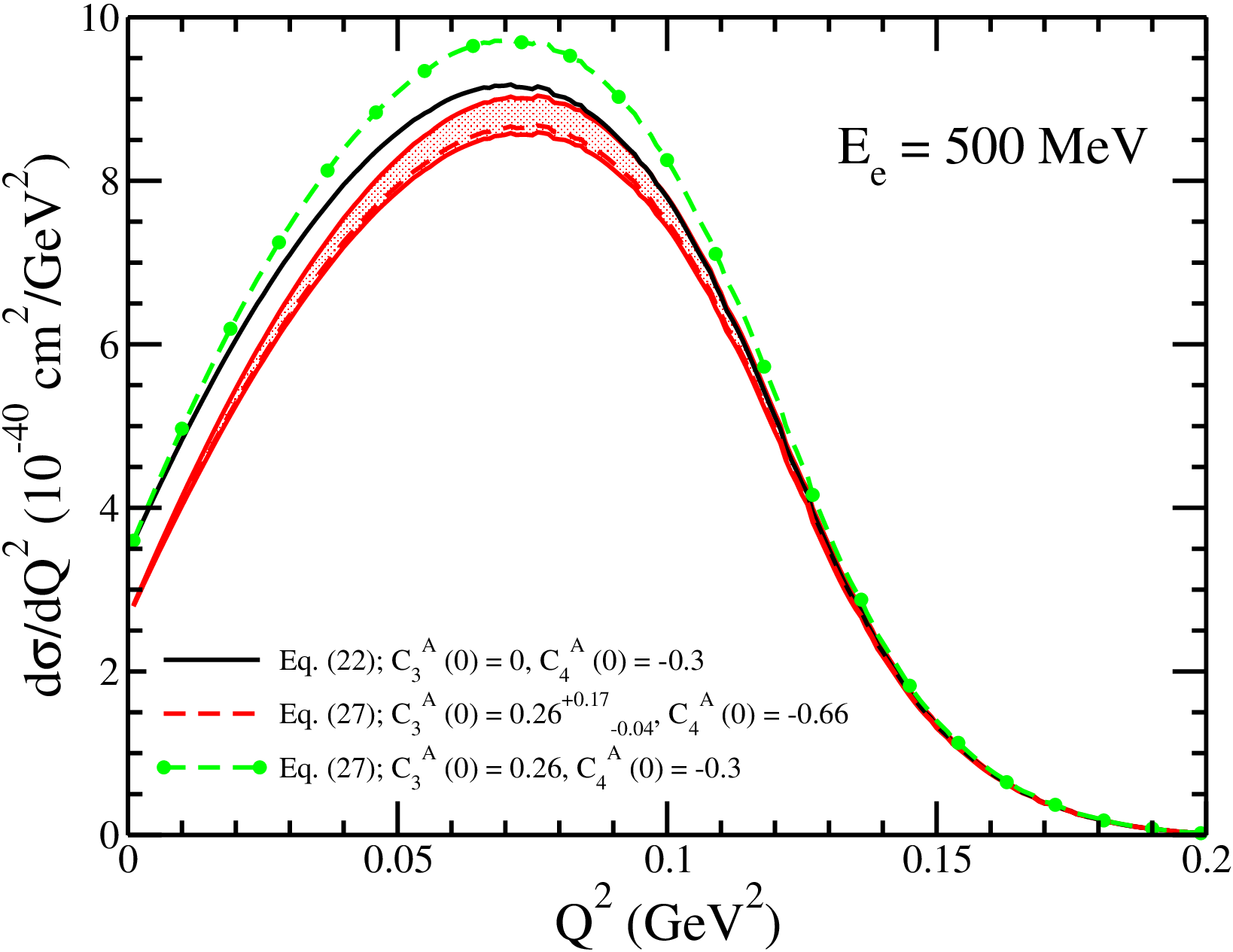}
\includegraphics[width=0.47\textwidth,height=.45\textwidth]{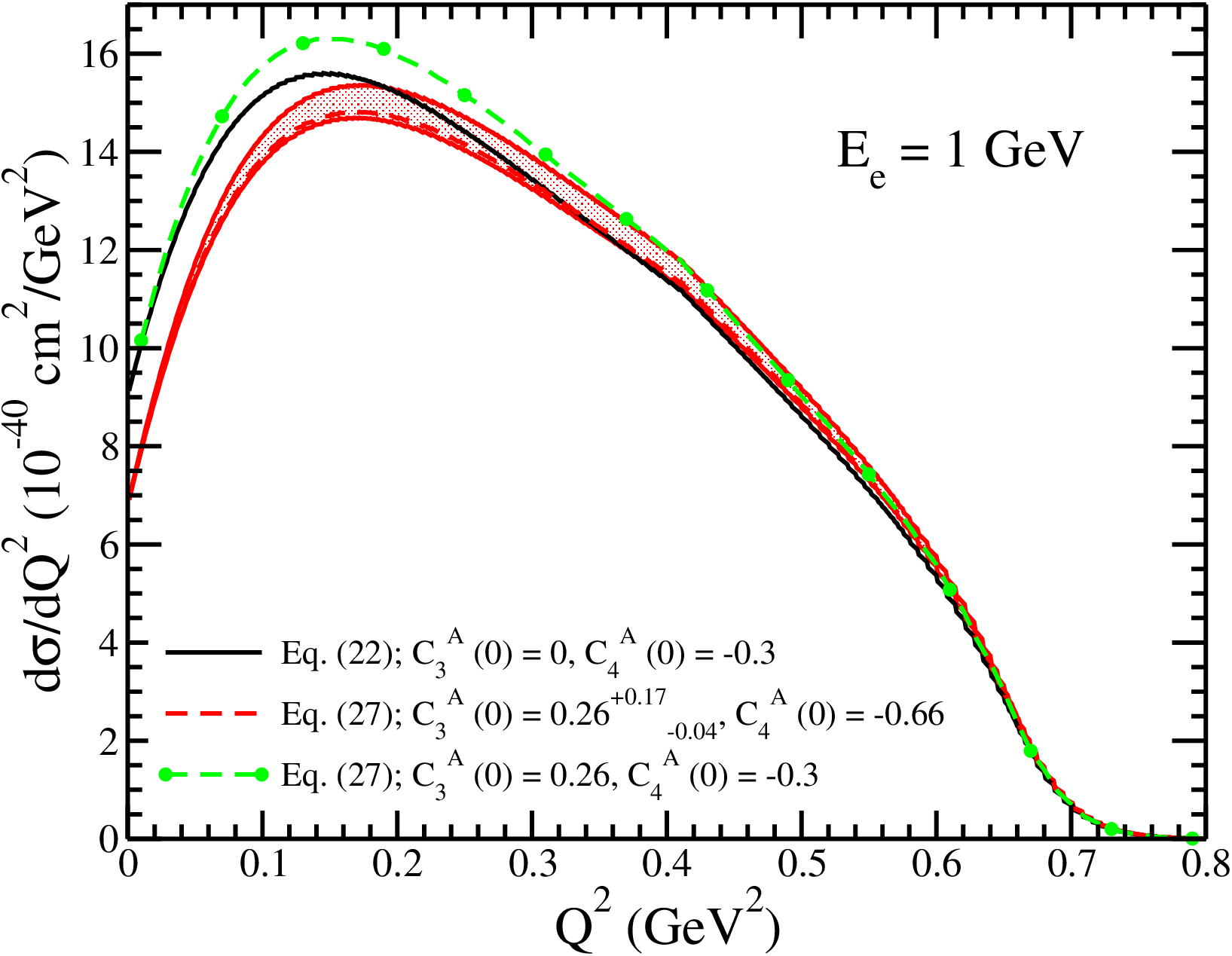}
\caption{$\frac{d\sigma}{dQ^2}$ as a function of $Q^2$ for the process $e^- + p \longrightarrow \nu_e + \Delta^0$ at $E_{e}=500$~MeV~(left panel) and 1~GeV~(right panel). 
The solid line shows the results obtained in the Adler's model~\cite{Adler:1968tw} using $C_{3}^{A}(0)=0$ and $C_{4}^{A}(0)=-0.3$. 
The dashed line~(dashed line with circle) corresponds to the result obtained in the model of Chen et al.~\cite{Chen:2023zhh} when $C_{3}^{A}(0)=0.26$ and $C_{4}^{A}(0)=-0.66$~($C_{3}^{A}(0)=0.26$ and $C_{4}^{A}(0)=-0.3$) is used. 
The band on the dashed line corresponds to the maximum and minimum allowed values for $C_{3}^{A}(0)$ i.e. $C_{3}^{A}(0)=0.26^{0.17}_{-0.04}$.
}
\label{fg:dsigma_nu_weak_electron_C3band}
\end{center}
\end{figure}

\begin{figure}
\begin{center}
\includegraphics[width=0.47\textwidth,height=.45\textwidth]{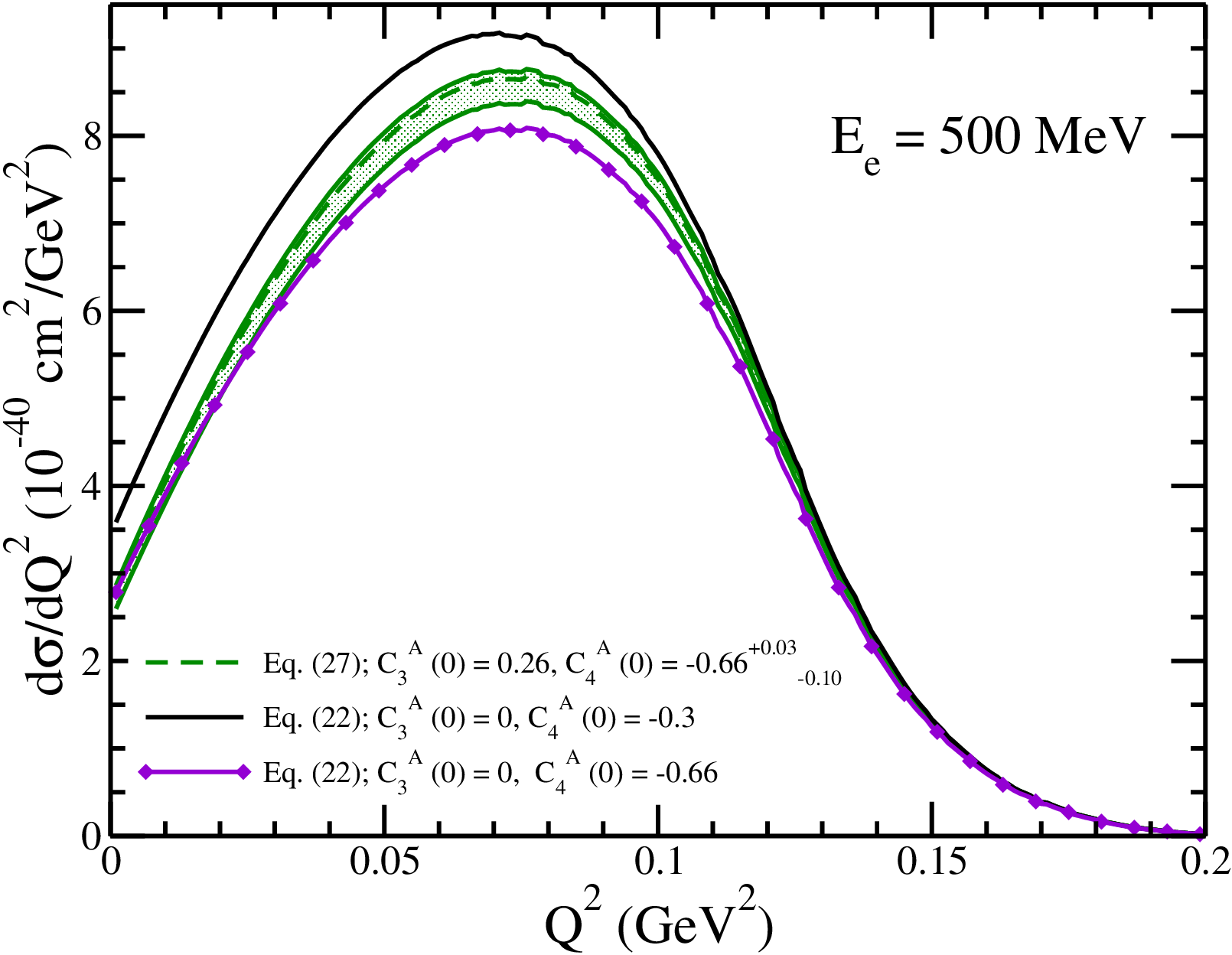}
\includegraphics[width=0.47\textwidth,height=.45\textwidth]{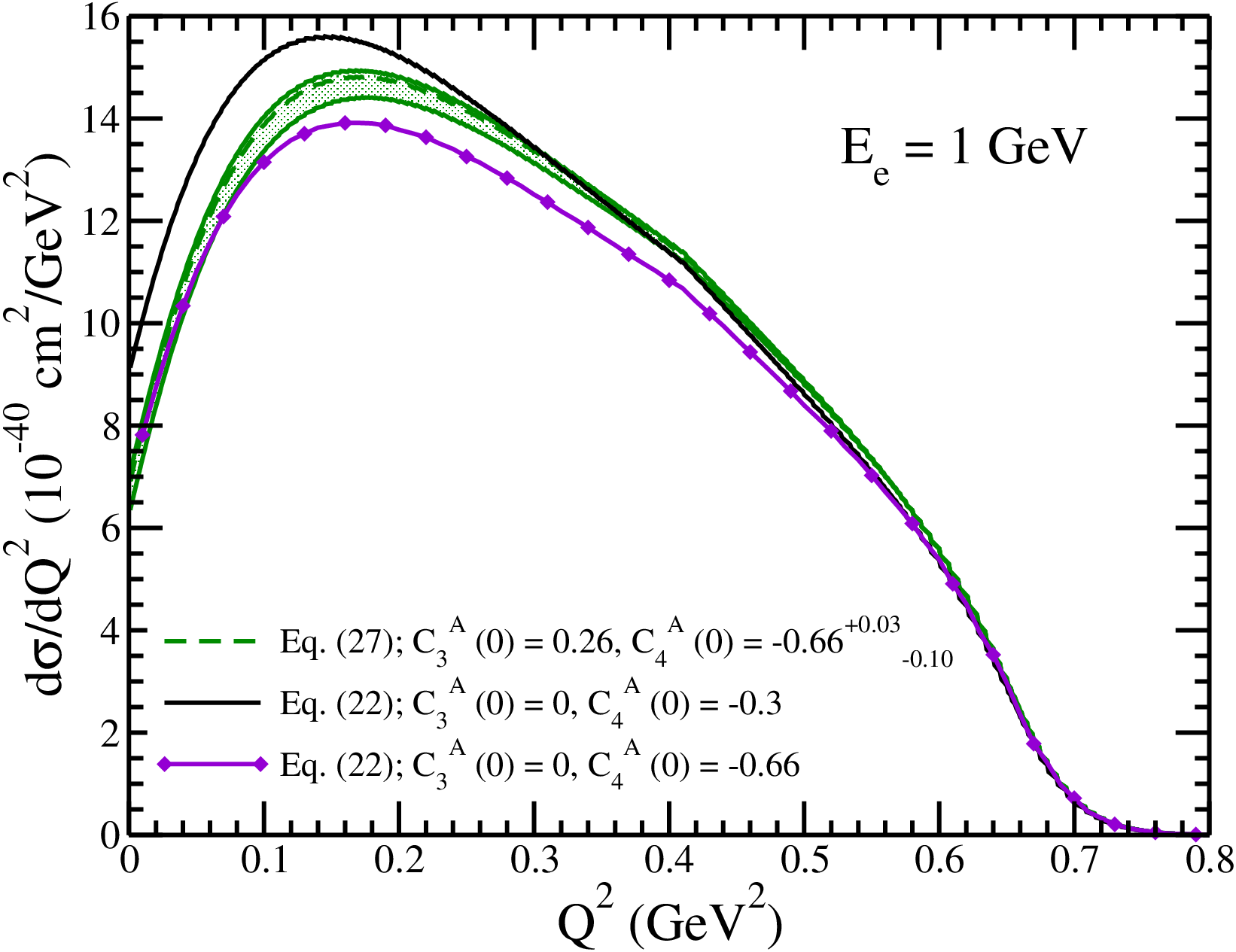}
\includegraphics[width=0.47\textwidth,height=.45\textwidth]{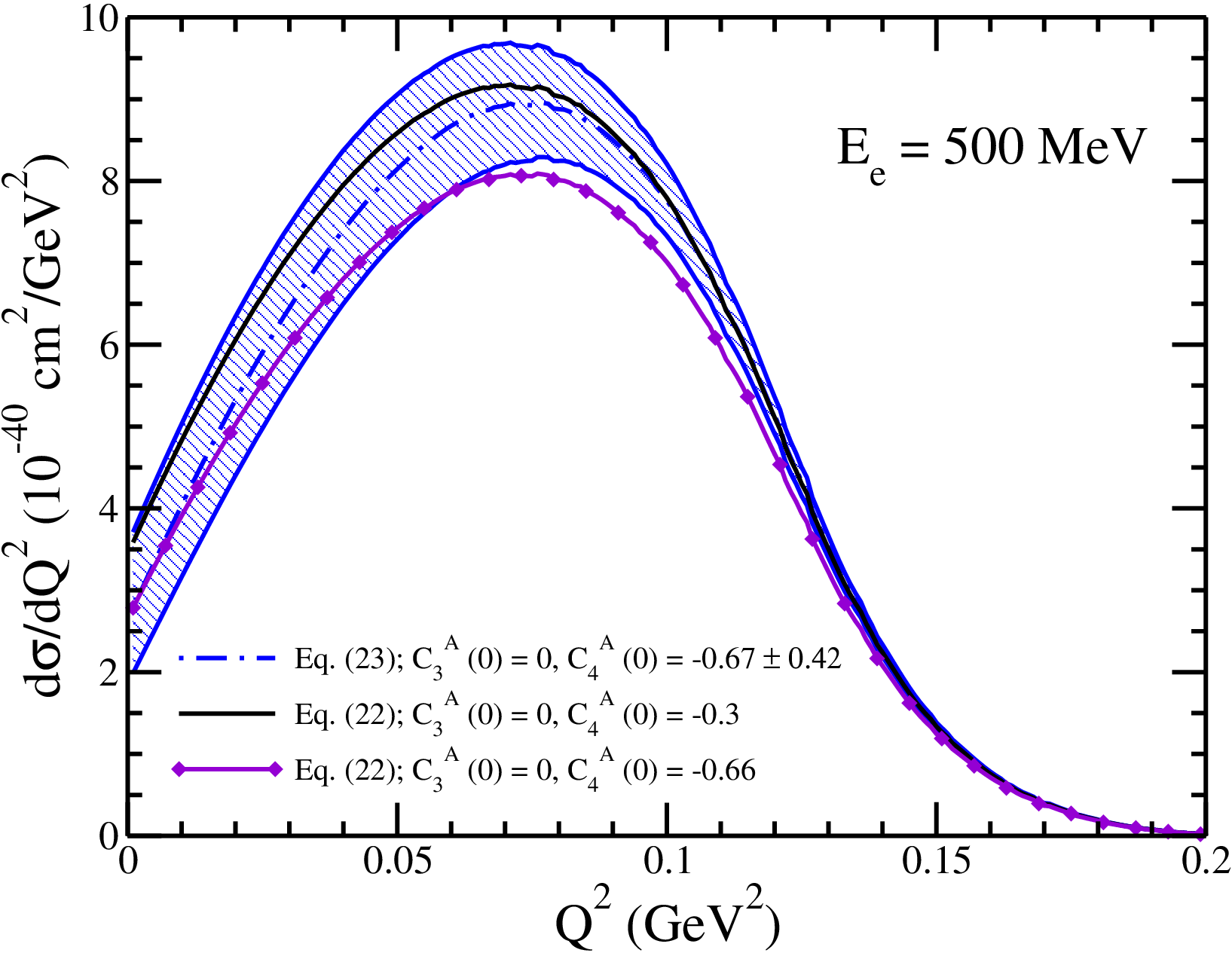}
\includegraphics[width=0.47\textwidth,height=.45\textwidth]{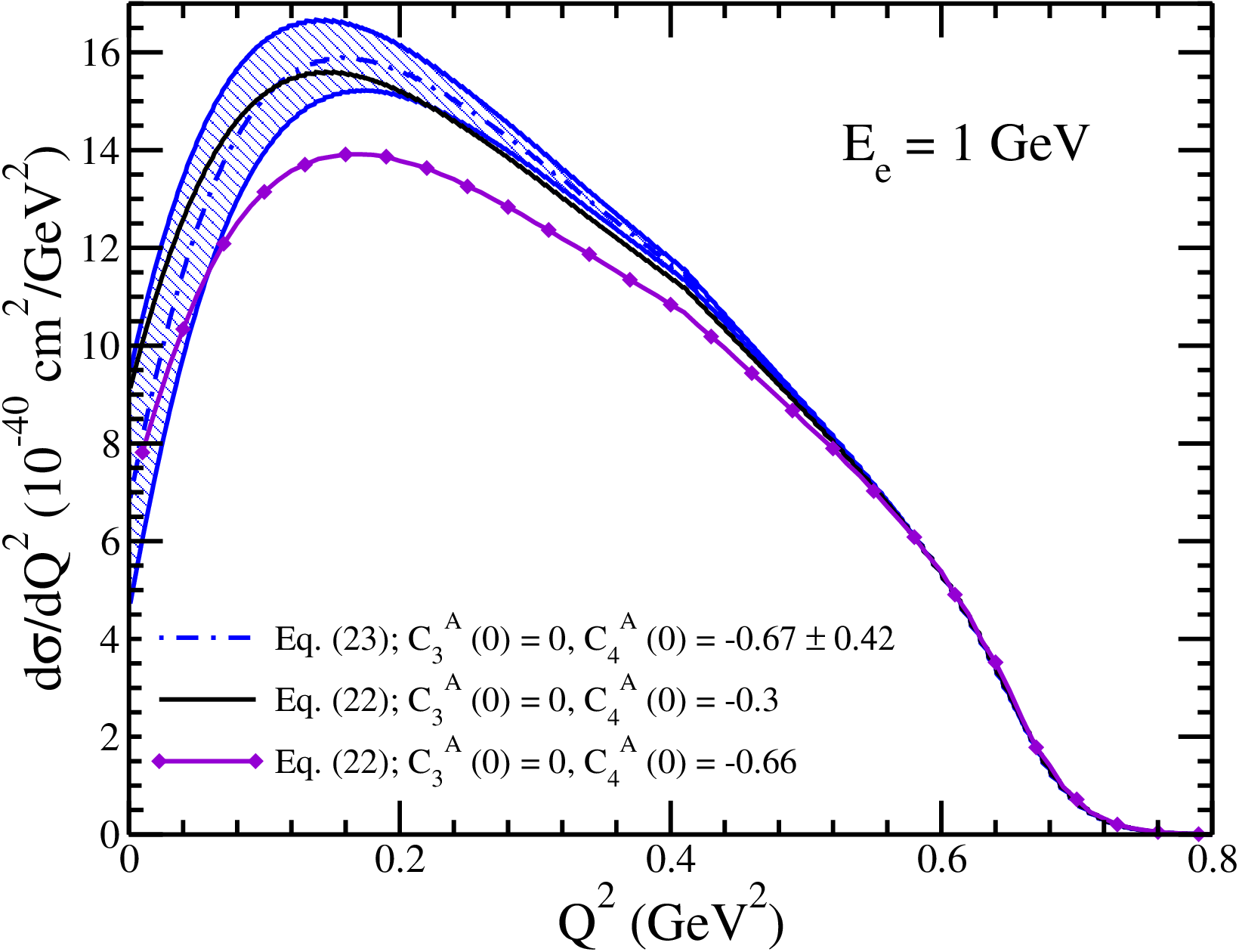}
\caption{$\frac{d\sigma}{dQ^2}$ as a function of $Q^2$ for the process $e^- + p \longrightarrow \nu_e + \Delta^0$ at $E_{e}=500$~MeV~(left panel) and 1~GeV~(right panel). 
The solid line~(solid line with diamonds) corresponds to the results obtained in the Adler's model~\cite{Adler:1968tw} using $C_{3}^{A}(0)=0$ and $C_{4}^{A}(0)=-0.3$~($C_{3}^{A}(0)=0$ and $C_{4}^{A}(0)=-0.66$). 
The dashed line corresponds to the results obtained in the model of Chen et al.~\cite{Chen:2023zhh} using $C_{3}^{A}(0)=0.26$ and $C_{4}^{A}(0)=-0.66$ and 
the band on the dashed line corresponds to the maximum and minimum allowed values of  $C_{4}^{A}(0)$ i.e. $C_{4}^{A} (0) = -0.66^{+0.03}_{-0.10}$. 
The dashed-dotted line corresponds to the results obtained using the model of Graczyk et al.~\cite{Graczyk:2009zh} with $C_{3}^{A}(0)=0$ and $C_{4}^{A}(0)=-0.67$ and the band corresponding to the dashed-dotted line is for $C_{4}^{A}(0)=-0.67 \pm 0.42$.}
\label{fg:dsigma_nu_weak_electron_C4band}
\end{center}
\end{figure}

It may be observed from the figure that the results obtained in the model of Barquilla-Cano et al.~\cite{Barquilla-Cano:2007vds} and Adler~\cite{Adler:1968tw} overlap with one another and the effect of the non-zero
value of $C_{3}^{A} (Q^2)$ parameterized using the theoretical model of Barquilla-Cano et al.~\cite{Barquilla-Cano:2007vds} is almost negligible for both the electron and positron induced reaction channels because of the very small value of $C_{3}^{A}(0) = 0.035$. 
However, the non-zero value of $C_{3}^{A} (Q^2)$ calculated in the model of Chen et al.~\cite{Chen:2023zhh} leads to an enhancement in the total scattering cross section for both the reaction channels. 
Quantitatively, in the case of electron induced reaction, the cross section increases by about 5\%~(8\%) when a value of $C_{3}^{A}(0) = 0.26~(0.4)$ is used as compared to the results obtained with $C_{3}^{A} (Q^2)=0$.
This increase in the cross section is almost the same in  the energy range of $E_{e}=1-2$~GeV, considered in this work. 
Contrary to the electron induced reaction, in the case of $\Delta^{++}$ production, at lower positron energy~($E_{e^{+}} \sim 1$~GeV) there is almost no effect of the non-zero value of $C_{3}^{A} (Q^2)$.
However, with the increase in the positron energy, the cross section obtained using $C_{3}^{A}(0) = 0.26~(0.4)$ increases as compared to the results obtained with $C_{3}^{A} (Q^2)=0$, and becomes 3\%~(6\%) at $E_{e^{+}} = 2$~GeV, which further increases to  5\%~(10\%) at $E_{e^{+}} = 4$~GeV.

\begin{figure}
\begin{center}
\includegraphics[width=0.47\textwidth,height=.45\textwidth]{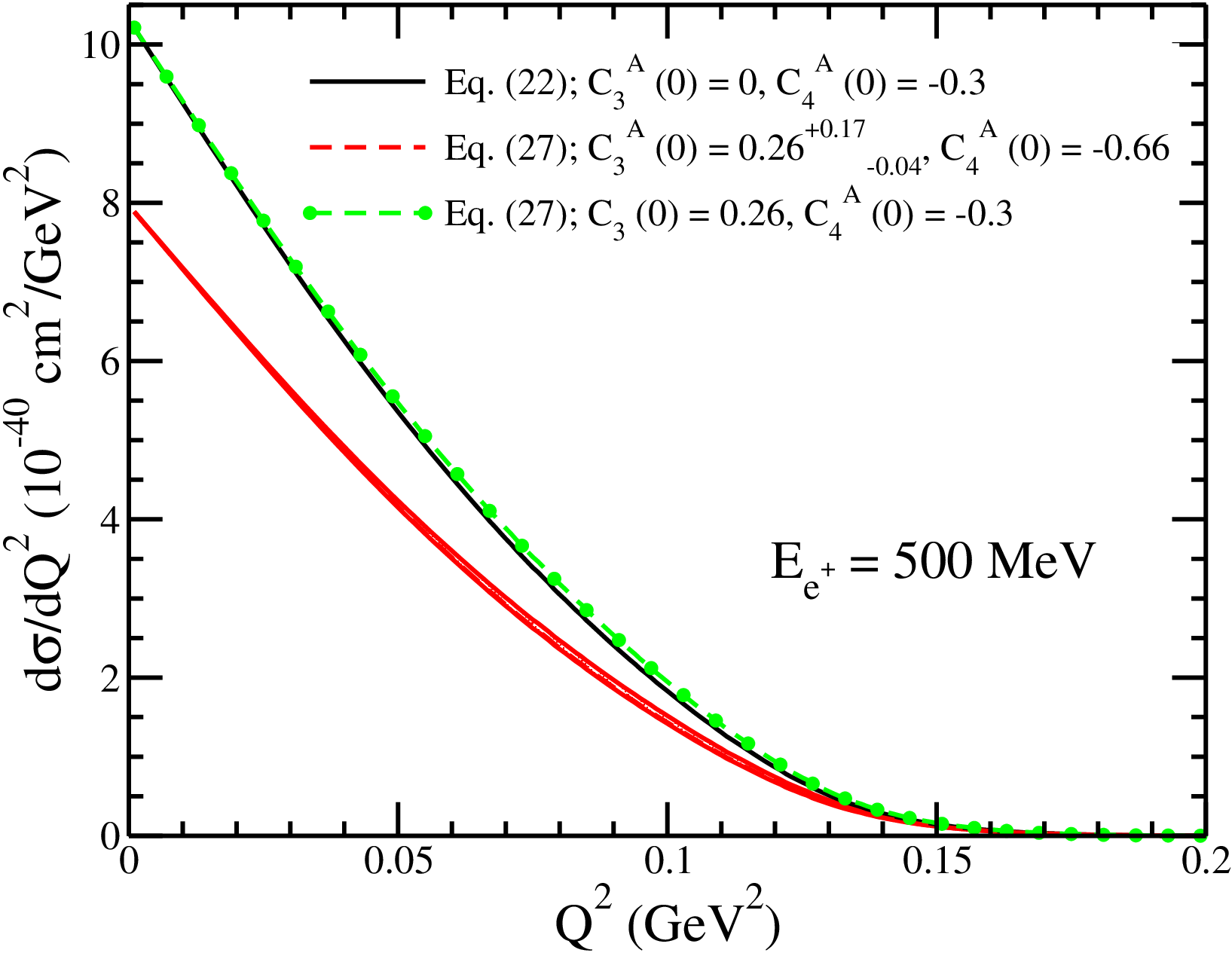}
\includegraphics[width=0.47\textwidth,height=.45\textwidth]{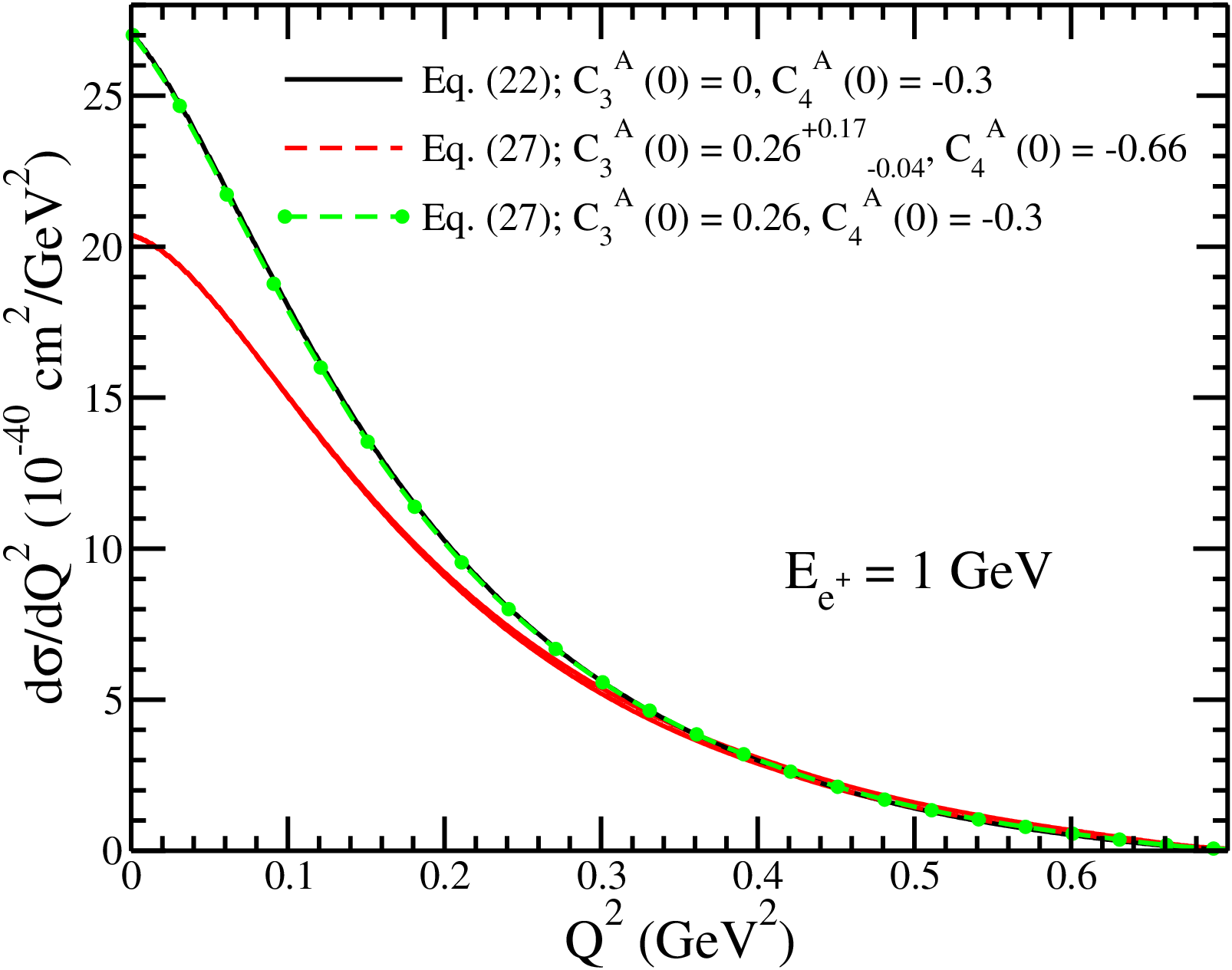}
\includegraphics[width=0.47\textwidth,height=.45\textwidth]{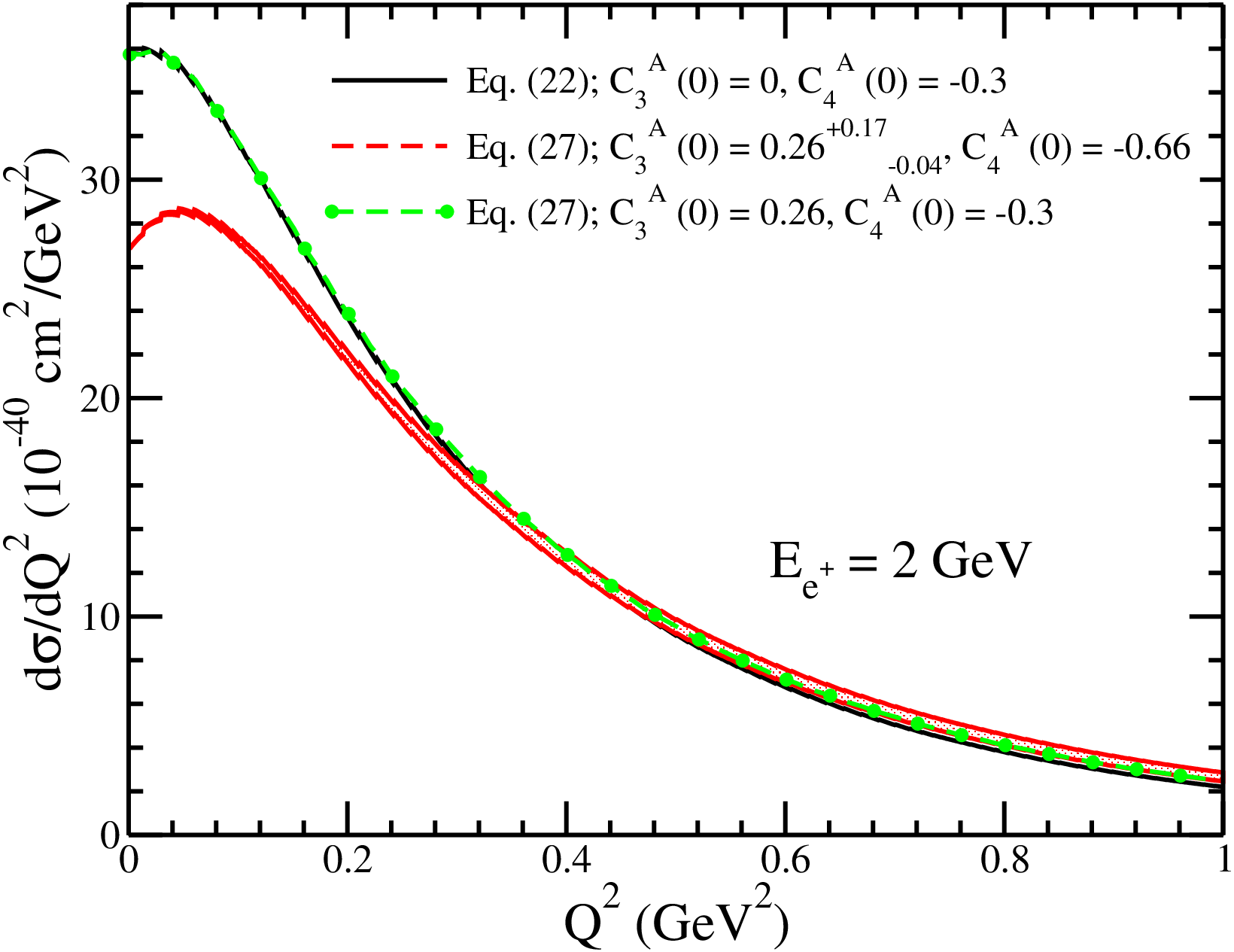}
\includegraphics[width=0.47\textwidth,height=.45\textwidth]{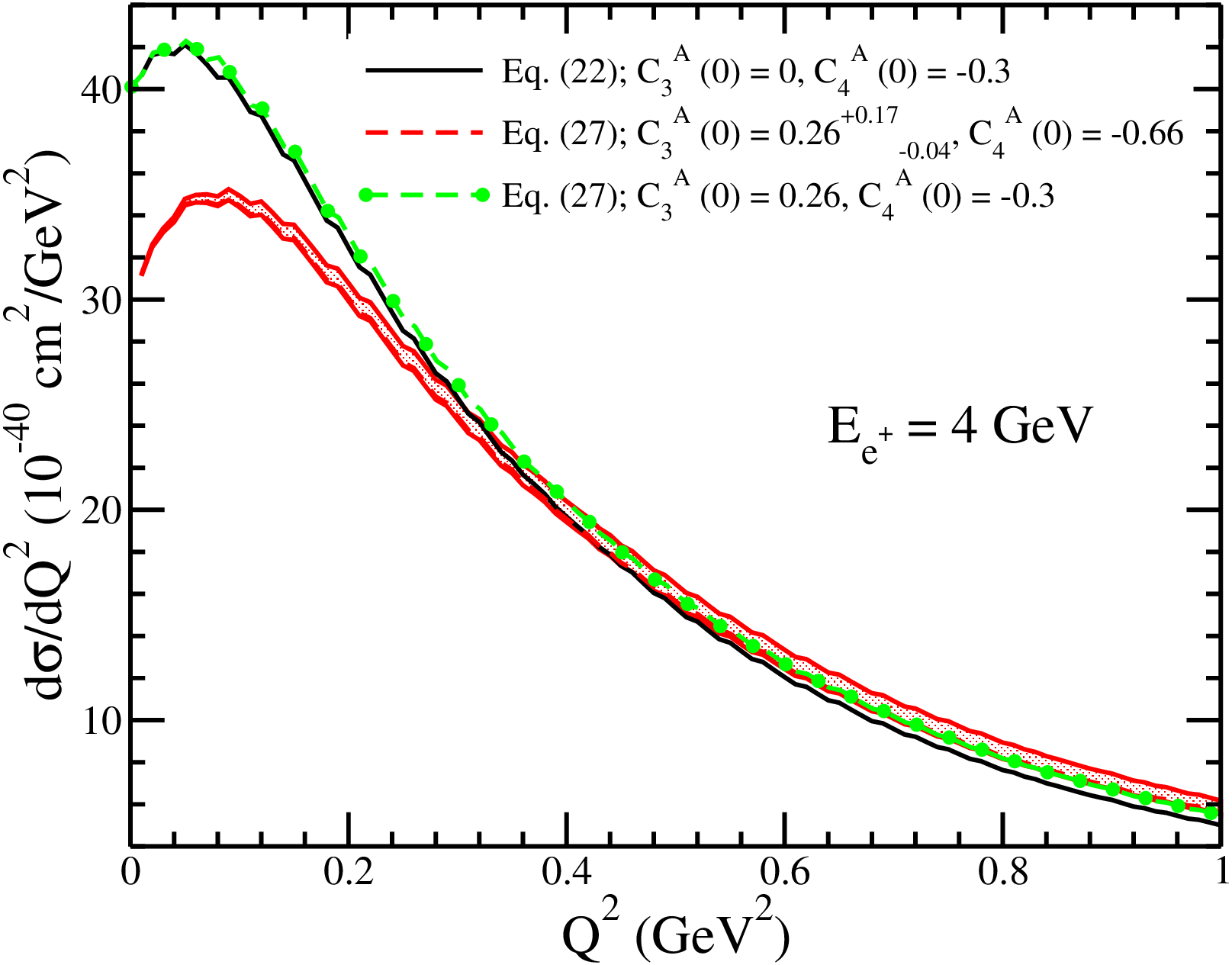}
\caption{$\frac{d\sigma}{dQ^2}$ as a function of $Q^2$ for the process $e^+ + p \longrightarrow \bar\nu_e + \Delta^{++}$ at $E_{e^{+}}=500$~MeV~(upper left panel),  1~GeV~(upper right panel), 2~GeV~(lower left panel) and 4~GeV~(lower right panel). Lines and points have the same meaning as in Fig.~\ref{fg:dsigma_nu_weak_electron_C3band}.}
\label{fg:dsigma_nu_weak_positron_C3band}
\end{center}
\end{figure}

To study the effect of $C_{4}^{A}(Q^2)$ on the total cross section, we obtain the results by taking different values of $C_{4}^{A}(0)$ as suggested by the model of Chen et al.~\cite{Chen:2023zhh} and Graczyk et al.~\cite{Graczyk:2009zh}. 
A smaller value of $C_{4}^{A} (0) = -0.66$ calculated in the model of Chen et al.~\cite{Chen:2023zhh}, with $C_{3}^{A}(0)=0.26$, leads to a reduction in the total cross section, which is more pronounced in the case of positron induced reaction as compared to the electron induced reaction. 
An interesting point to mention here is the comparison of the results calculated using the values of $C_{3}^{A}(0)=0$ and $C_{4}^{A} (0) =-0.3$ in the Adler's model~\cite{Adler:1968tw}  with the results obtained using the  central values in the model of Chen et al.~\cite{Chen:2023zhh} i.e. $C_{3}^{A}(0)=0.26$ and $C_{4}^{A} (0) =-0.66$. 
We find that, in the case of electron induced $\Delta$ production process, the results obtained in the two models are quite consistent with one another, 
while in the case of positron induced process, a suppression in the total cross section is observed in the model of Chen et al.~\cite{Chen:2023zhh} as compared to the results obtained in Adler's model~\cite{Adler:1968tw}. 

Quantitatively, this suppression is $\approx$ 20\% at $E_{e^{+}}=0.5$~GeV, which decreases with the increase in positron energy and becomes 9\% and 6\% at  $E_{e^{+}}=0.2$~GeV and 4~GeV, respectively. 
Furthermore, comparing the results obtained using Adler's model~\cite{Adler:1968tw} with the results obtained using the parameterization given by Graczyk~\cite{Graczyk:2009zh}~(Eq.~(\ref{C4;Graczyk}) with $C_{3}^{A}(0)=0$ and $C_{4}^{A} (0) = -0.67$), in the case of electron~(positron) induced processes,  we find a suppression in the cross section of about 5\%~(10\%) in the region of low energy, which becomes almost negligible at $E_{e} \gtrsim 1$~GeV.

Thus, we find that taking a non-zero value of $C_{3}^{A} (0)$ as suggested in some of the theoretical models and keeping $C_{4}^{A}(0)$ as predicted by the Adler's model~\cite{Adler:1968tw} results in the increase in the cross section, both for the electron and positron induced processes. 
While keeping the same value of $C_{3}^{A}(0)$ and using a smaller value of $C_{4}^{A}(0)$ results in a decrease in the cross section, which is more for the positron induced process as compared to the electron induced reactions.

 \subsection{Differential cross section $\frac{d\sigma}{dQ^2}$ for various energies of electron and positron}
\begin{figure}
\begin{center}
\includegraphics[width=0.47\textwidth,height=.45\textwidth]{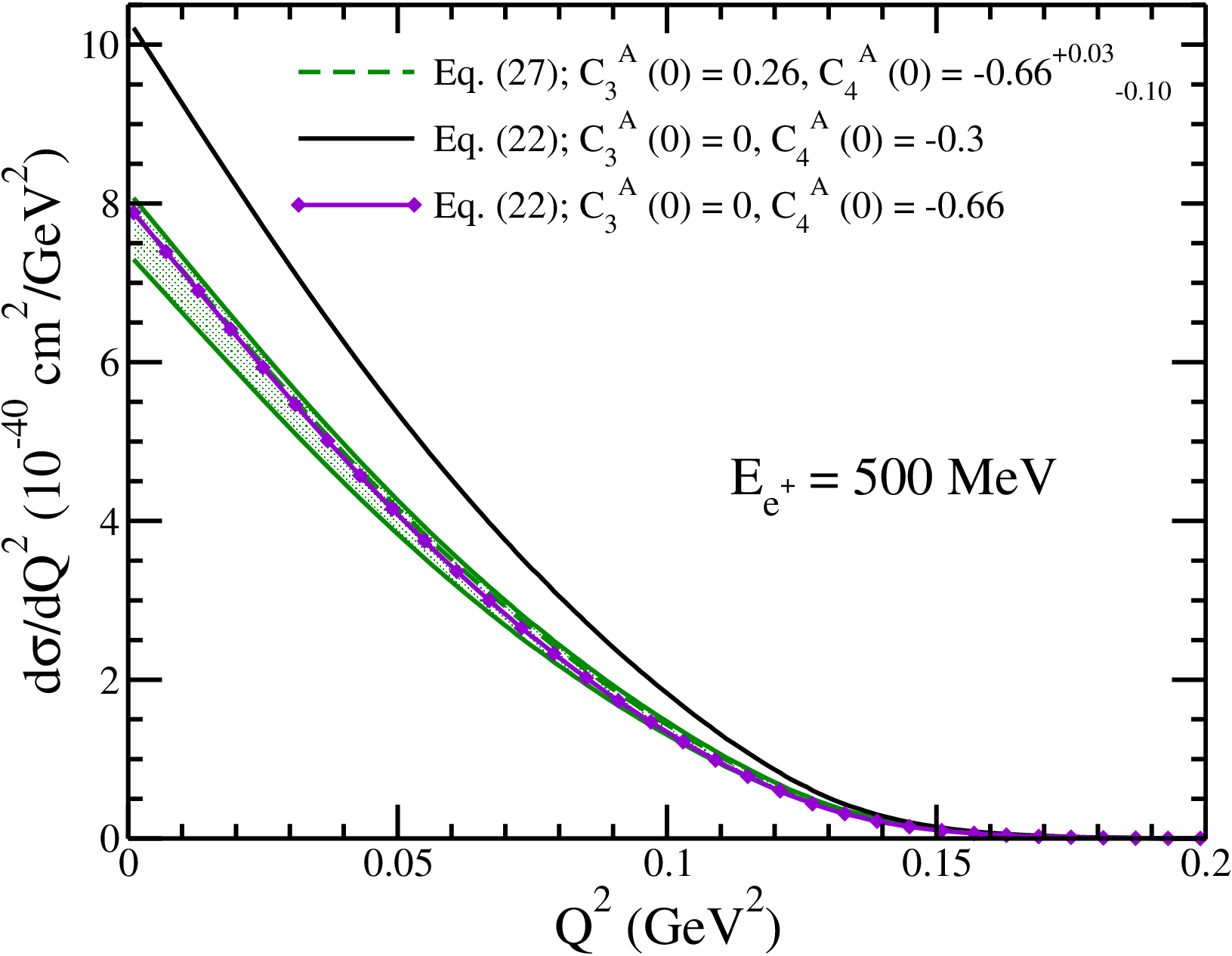}
\includegraphics[width=0.47\textwidth,height=.45\textwidth]{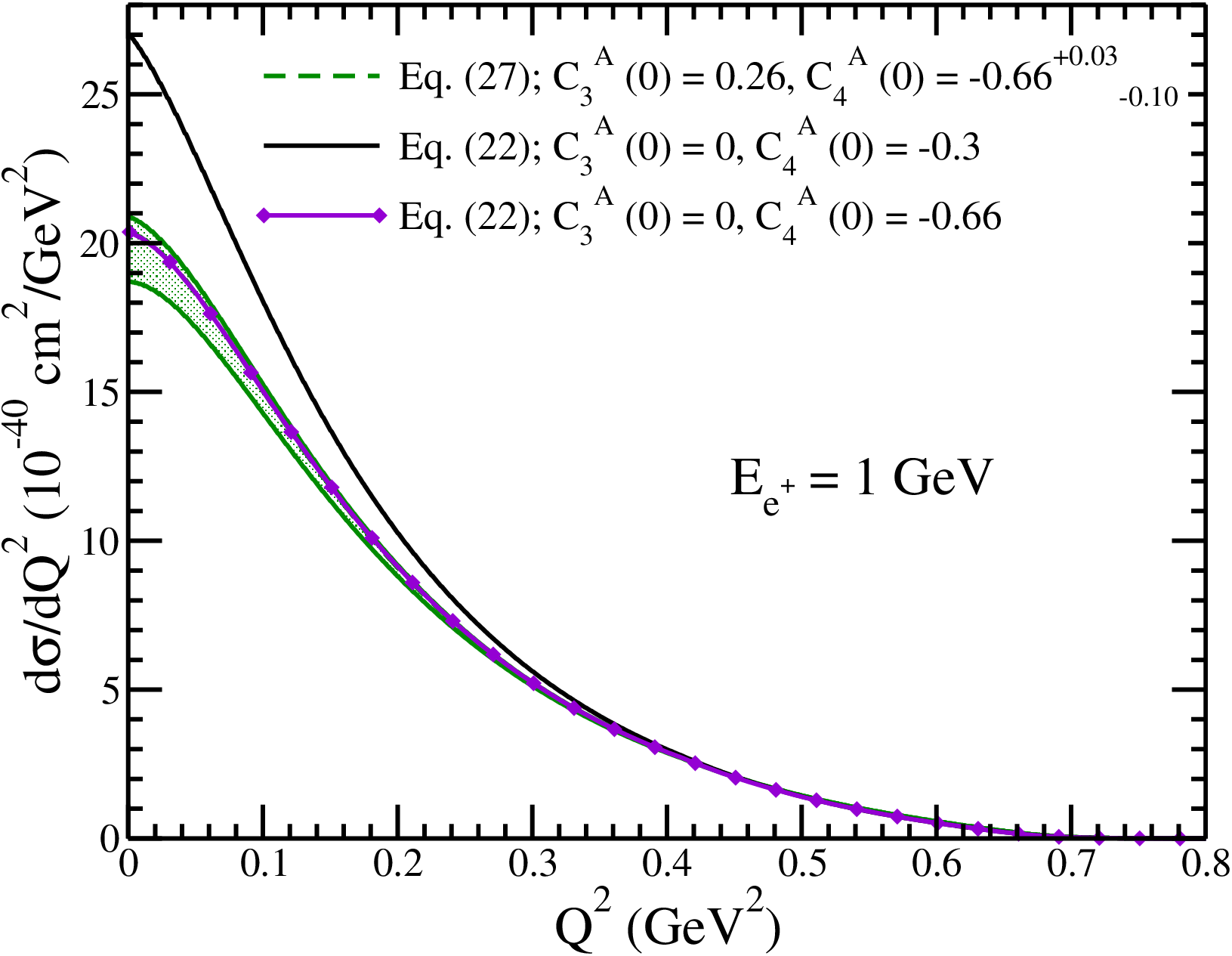}
\includegraphics[width=0.47\textwidth,height=.45\textwidth]{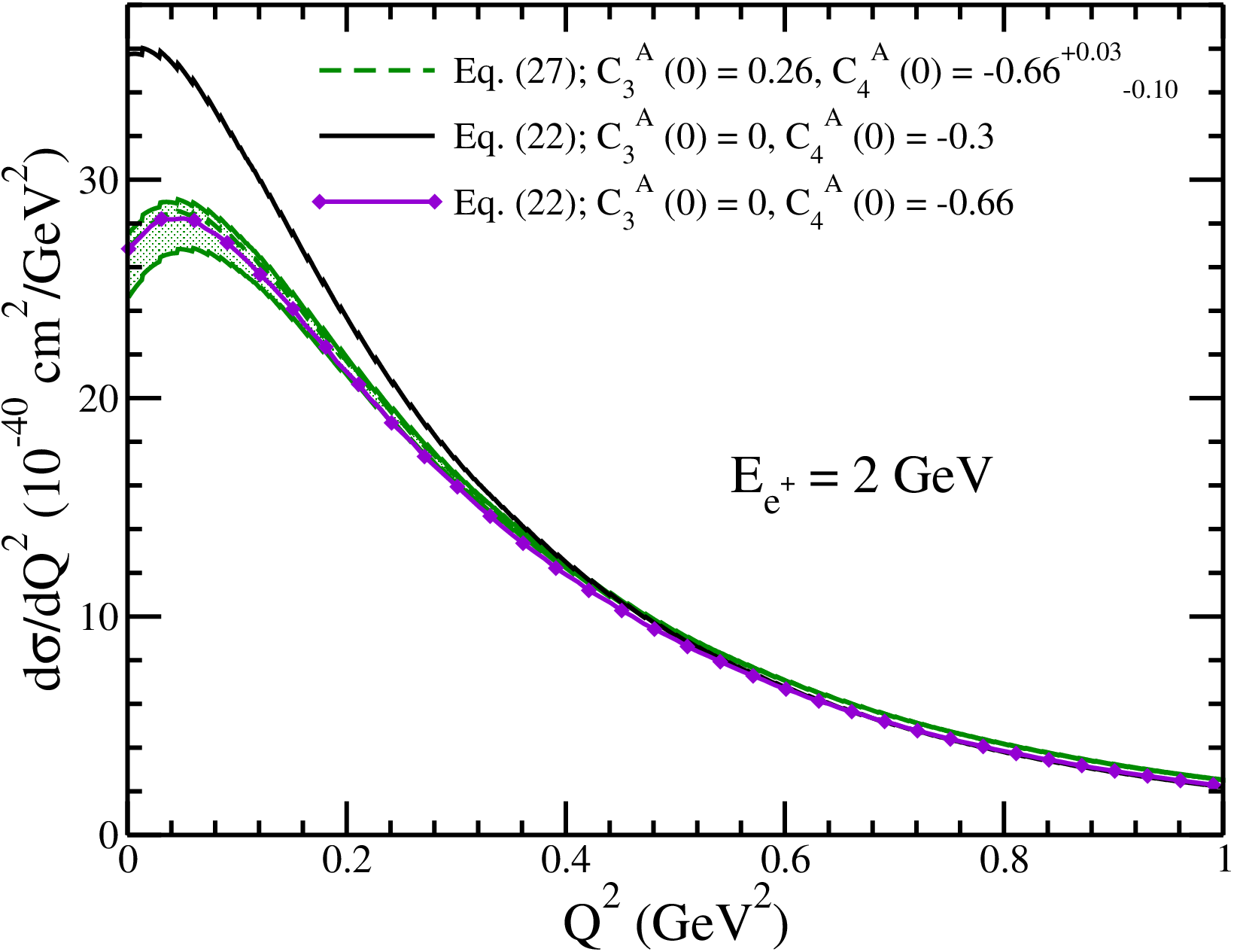}
\includegraphics[width=0.47\textwidth,height=.45\textwidth]{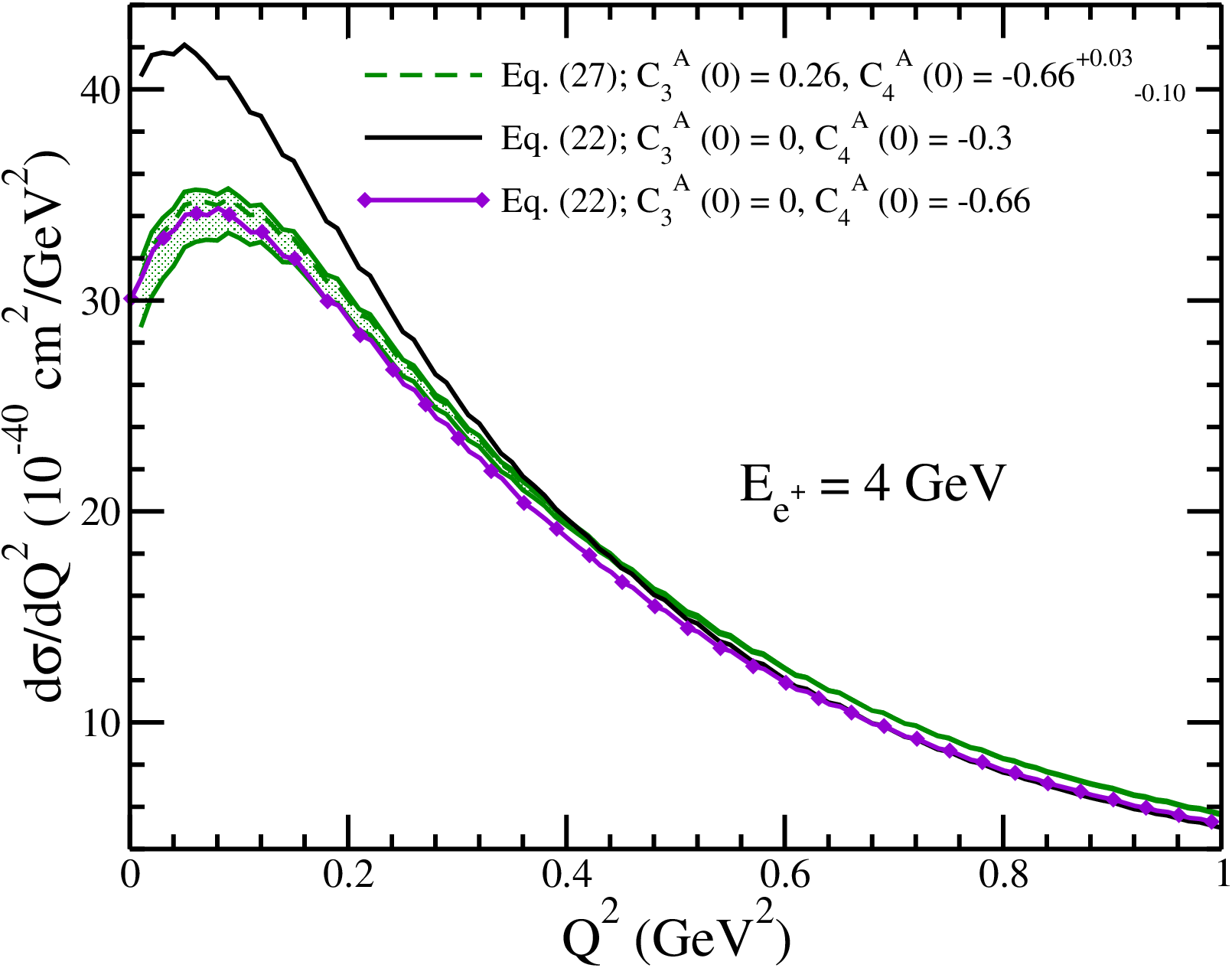}
\caption{$\frac{d\sigma}{dQ^2}$ as a function of $Q^2$ for the process $e^+ + p \longrightarrow \bar\nu_e + \Delta^{++}$ at $E_{e^{+}}=500$~MeV~(upper left panel),  1~GeV~(upper right panel), 2~GeV~(lower left panel) and 4~GeV~(lower right panel). Lines and points have the same meaning as in Fig.~\ref{fg:dsigma_nu_weak_electron_C4band}.}
\label{fg:dsigma_nu_weak_positron_C4band_Chen}
\end{center}
\end{figure}
\begin{figure}
\begin{center}
\includegraphics[width=0.47\textwidth,height=.45\textwidth]{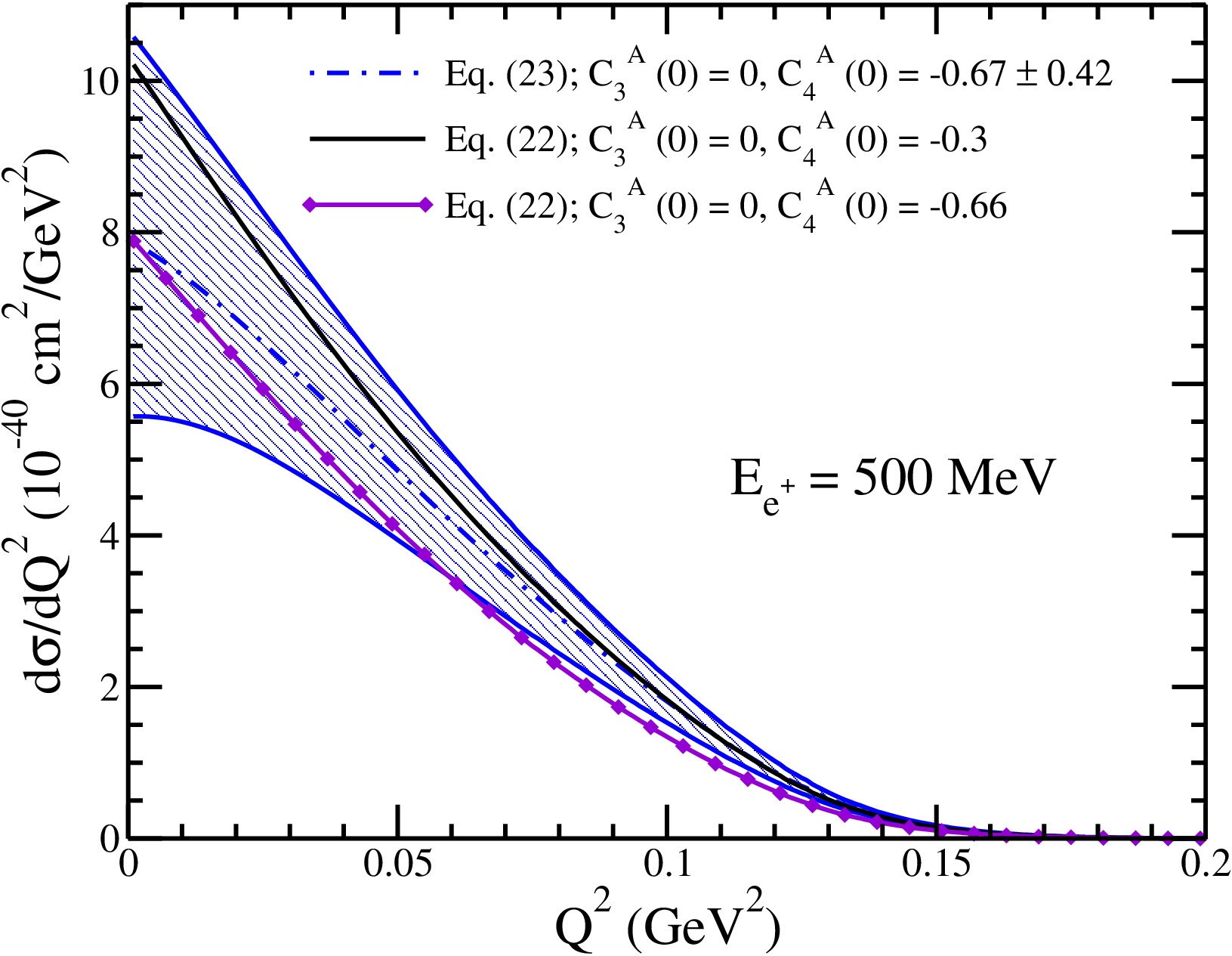}
\includegraphics[width=0.47\textwidth,height=.45\textwidth]{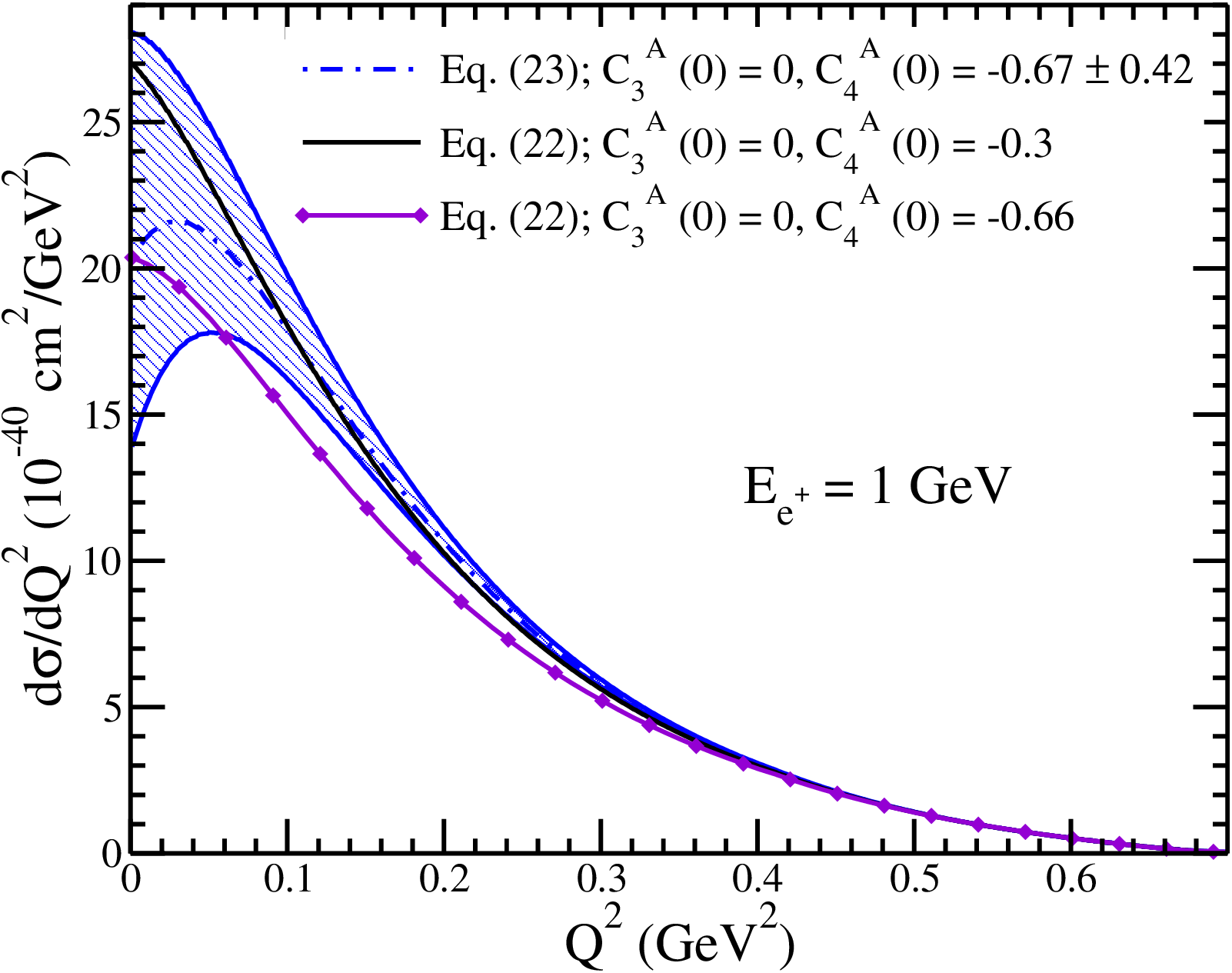}
\includegraphics[width=0.47\textwidth,height=.45\textwidth]{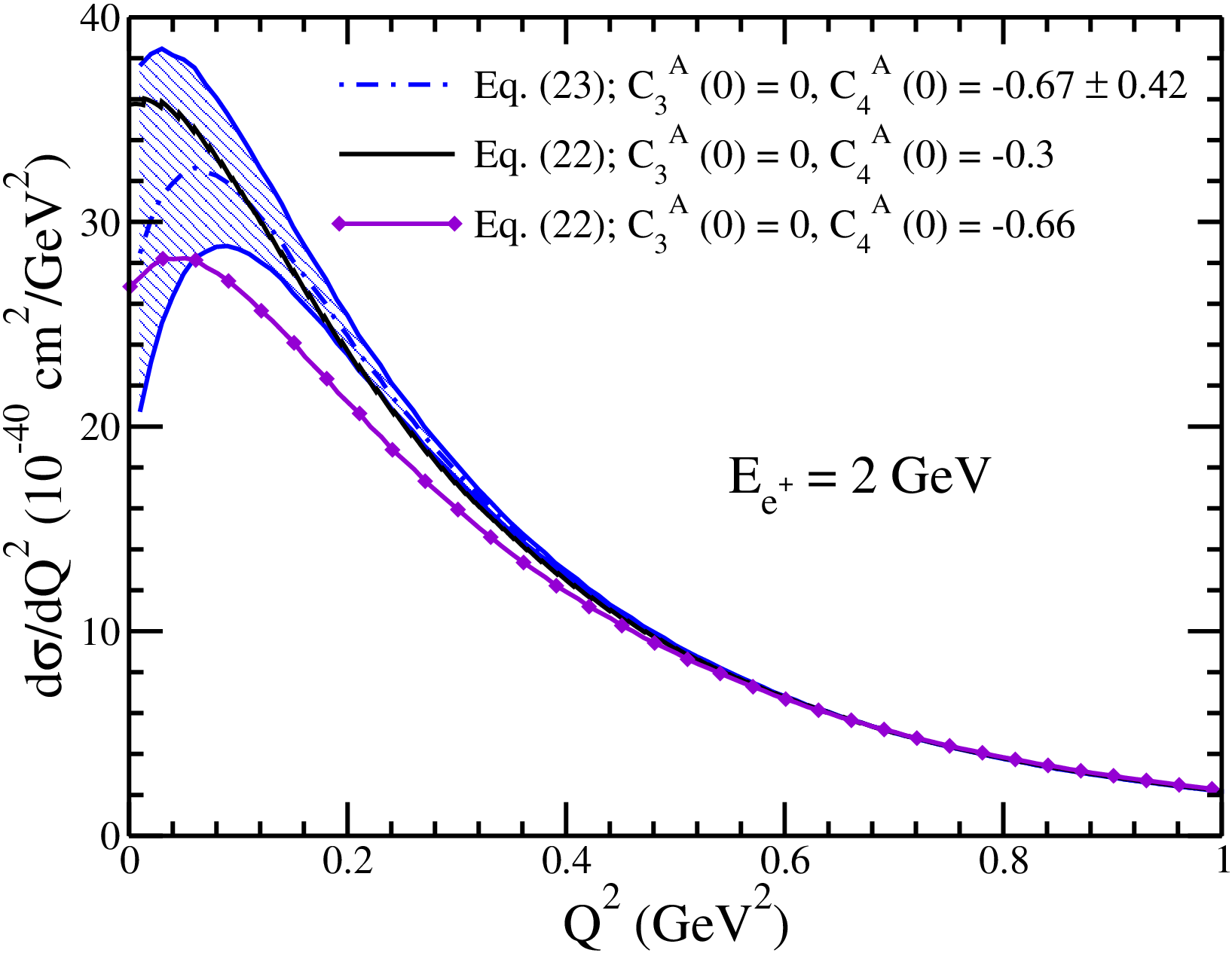}
\includegraphics[width=0.47\textwidth,height=.45\textwidth]{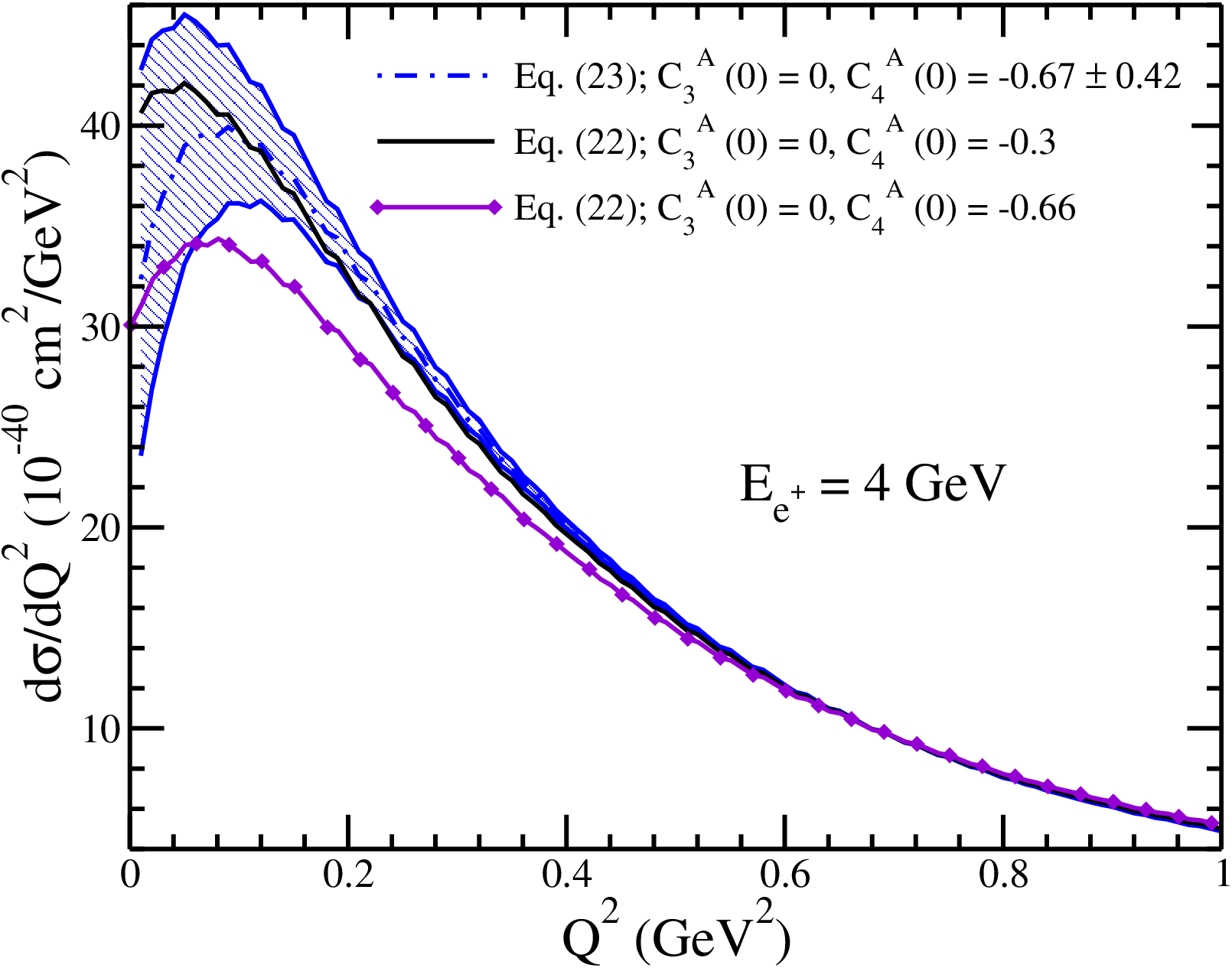}
\caption{$\frac{d\sigma}{dQ^2}$ as a function of $Q^2$ for the process $e^+ + p \longrightarrow \bar\nu_e + \Delta^{++}$ at $E_{e^{+}}=500$~MeV~(upper left panel),  1~GeV~(upper right panel), 2~GeV~(lower left panel) and 4~GeV~(lower right panel). Lines and points have the same meaning as in Fig.~\ref{fg:dsigma_nu_weak_electron_C4band}.}
\label{fg:dsigma_nu_weak_positron_C4band_Graczyk}
\end{center}
\end{figure}

In Figs.~\ref{fg:dsigma_compare}--\ref{fg:dsigma_nu_weak_positron_C4band_Graczyk}, we present the results for $\frac{d\sigma}{dQ^2}$ vs. $Q^2$ for the processes $e^- + p \longrightarrow {\nu}_e + \Delta^{0}$ and $e^+ + p \longrightarrow \bar{\nu}_e + \Delta^{++}$ at various electron/positron energies. 
We have studied the differential cross section $\frac{d\sigma}{dQ^2}$ vs. $Q^2$ in order to explore the effect of $C_{3}^{A}(Q^2)$ and $C_{4}^{A}(Q^2)$ using the various parameterizations for these form factors while using 
the vector form factors $C_{i}^{V}(Q^2);~(i=3-5)$ and the axial vector form factor $C_{5}^{A} (Q^2)$ given in Eqs.~(\ref{civ_lala}) and (\ref{cia_lala}), respectively.

 In Fig.~\ref{fg:dsigma_compare}, we present the results of the $Q^2$ distribution for the processes $e^- + p \longrightarrow {\nu}_e + \Delta^{0}$ and $e^+ + p \longrightarrow \bar{\nu}_e + \Delta^{++}$ in the region of low $Q^2$~($Q^{2}<0.1$~GeV$^{2}$) at $E_{e}=0.7$~GeV, which corresponds broadly to the average neutrino energy of the ANL experiment, to highlight the dependence of $\frac{d\sigma}{dQ^2}$ on the different values of $C_{3}^{A}(Q^2)$ and $C_{4}^{A}(Q^2)$ which have been used in the analysis of the of the neutrino experiment of ANL by Barish et al.~\cite{Barish:1978pj} and in a later analysis by Graczyk et al.~\cite{Graczyk:2009qm, Graczyk:2009zh}, which prefer the value $C_{3}^{A}(0) =0$ and $C_{4}^{A} (0) = - \frac{1}{4} C_{5}^{A} (0)$ as suggested by the Adler's model~\cite{Adler:1968tw}. 
 We see from the figure that $\frac{d\sigma}{dQ^2}$ vs. $Q^2$, in the region of very low $Q^2$ i.e. $0<Q^2 <0.1$~GeV$^{2}$, is more sensitive to the values taken for the form factors $C_{3}^{A}(0)$ and $C_{4}^{A} (0)$ as compared to the neutrino production of $\Delta^{++}$, as observed in Fig.~17(b) of Barish et al.~\cite{Barish:1978pj} in the case of the neutrino experiment. 
 It should also be emphasized from the results of the present work that this sensitivity of $\frac{d\sigma}{dQ^2}$ on the choice of the values of the form factors $C_{3}^{A}(0)$ and $C_{4}^{A} (0)$ is quite large in the case of positron induced $\Delta^{++}$ production as compared to the electron induced $\Delta^{0}$ production. 
 An observation of $\frac{d\sigma}{dQ^2}$ in the low $Q^2$ region in the case of positron produced weak $\Delta^{++}$ production could, therefore, be decisive in determining, phenomenologically, the values of $C_{3}^{A}(0)$ and $C_{4}^{A} (0)$.

\begin{figure}
\begin{center}
\includegraphics[width=0.47\textwidth,height=.45\textwidth]{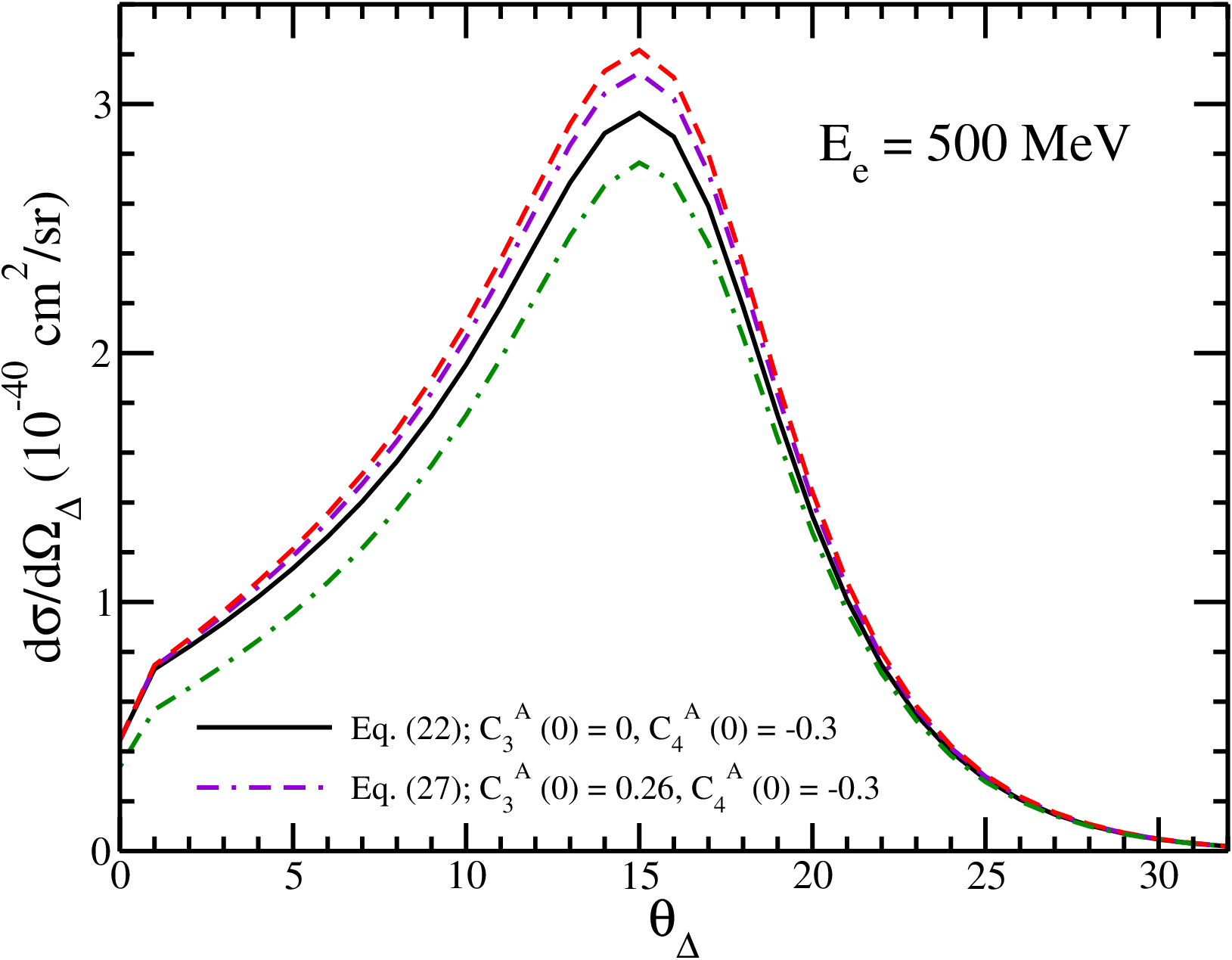}
\includegraphics[width=0.47\textwidth,height=.45\textwidth]{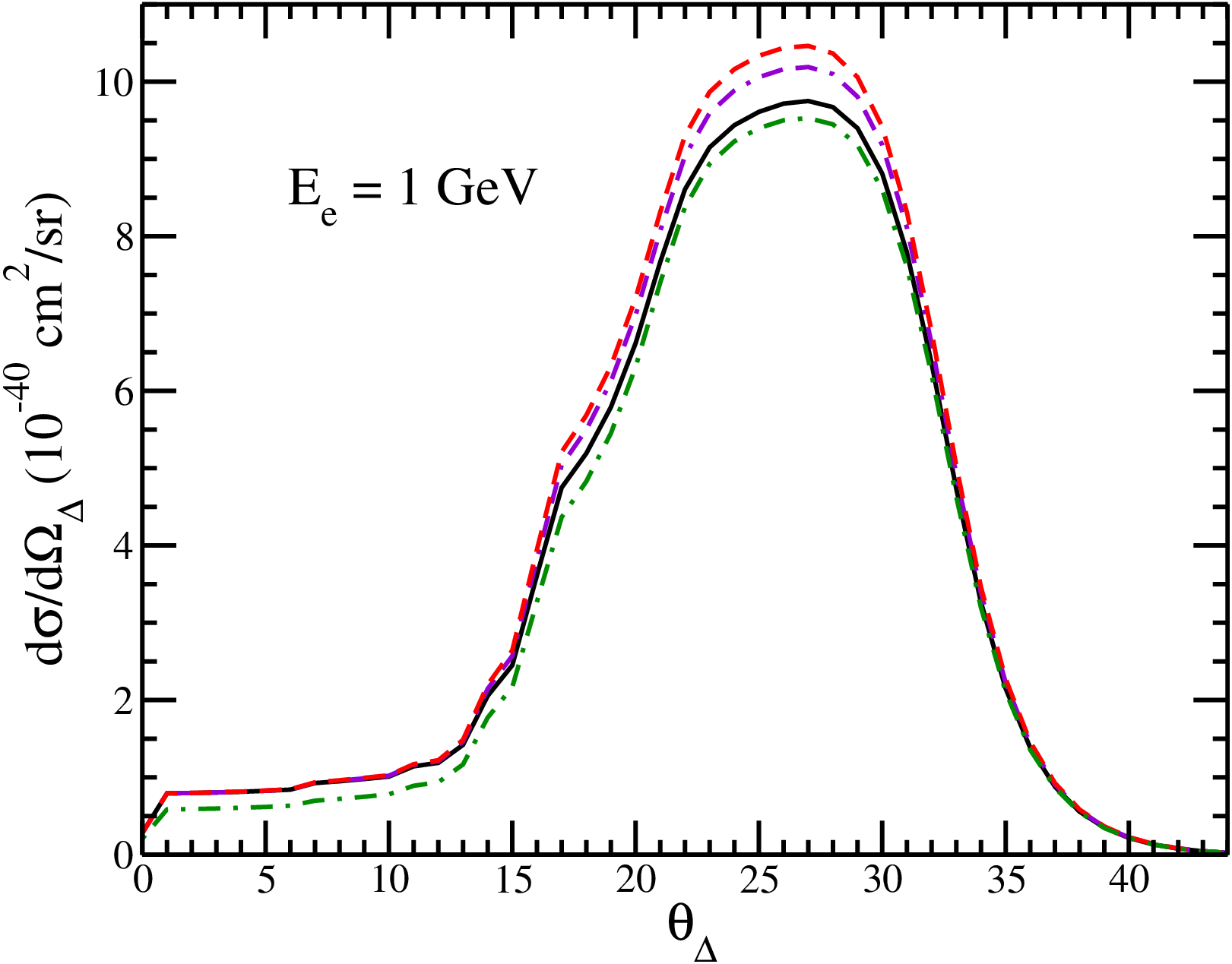}
\includegraphics[width=0.47\textwidth,height=.45\textwidth]{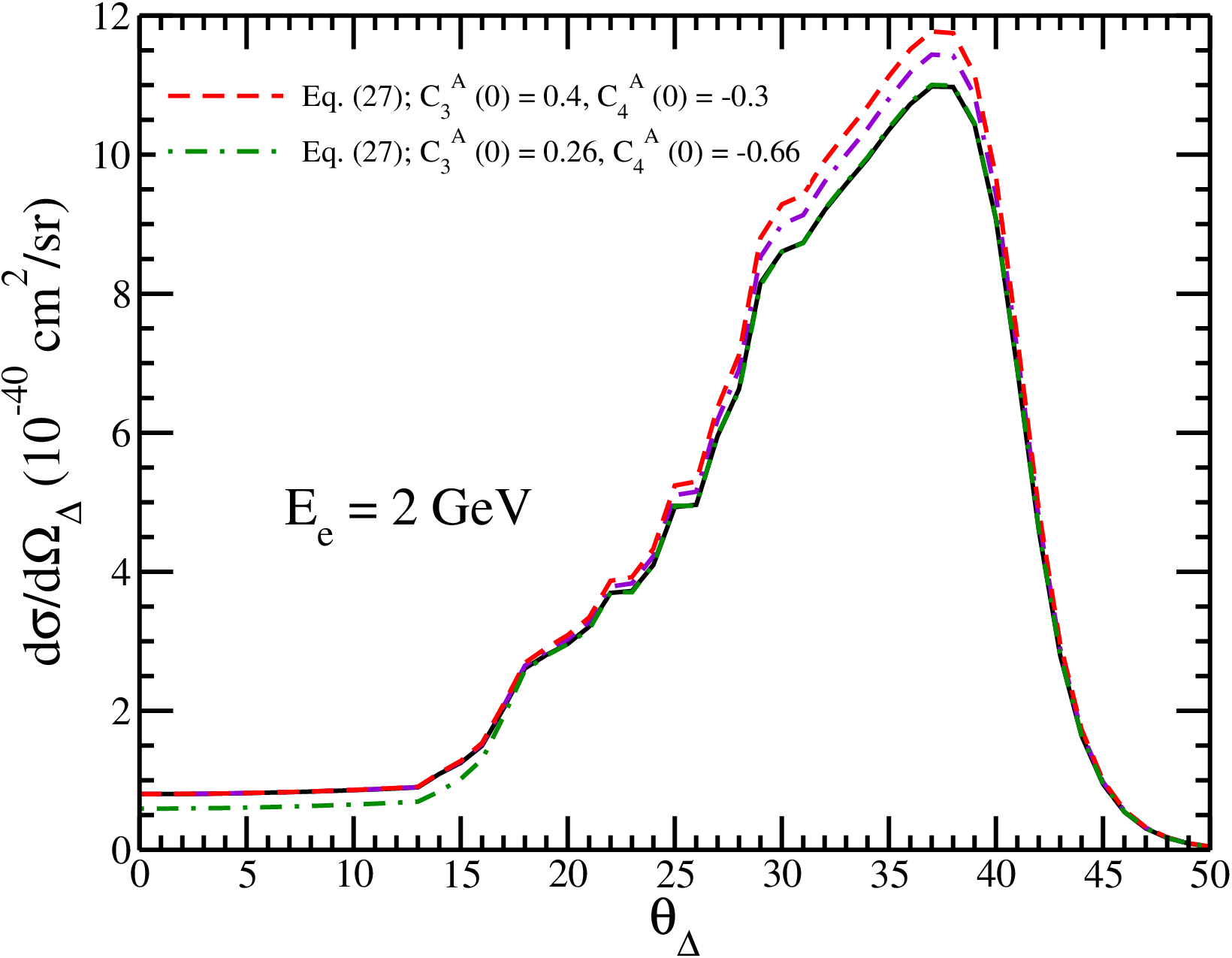}
\includegraphics[width=0.47\textwidth,height=.45\textwidth]{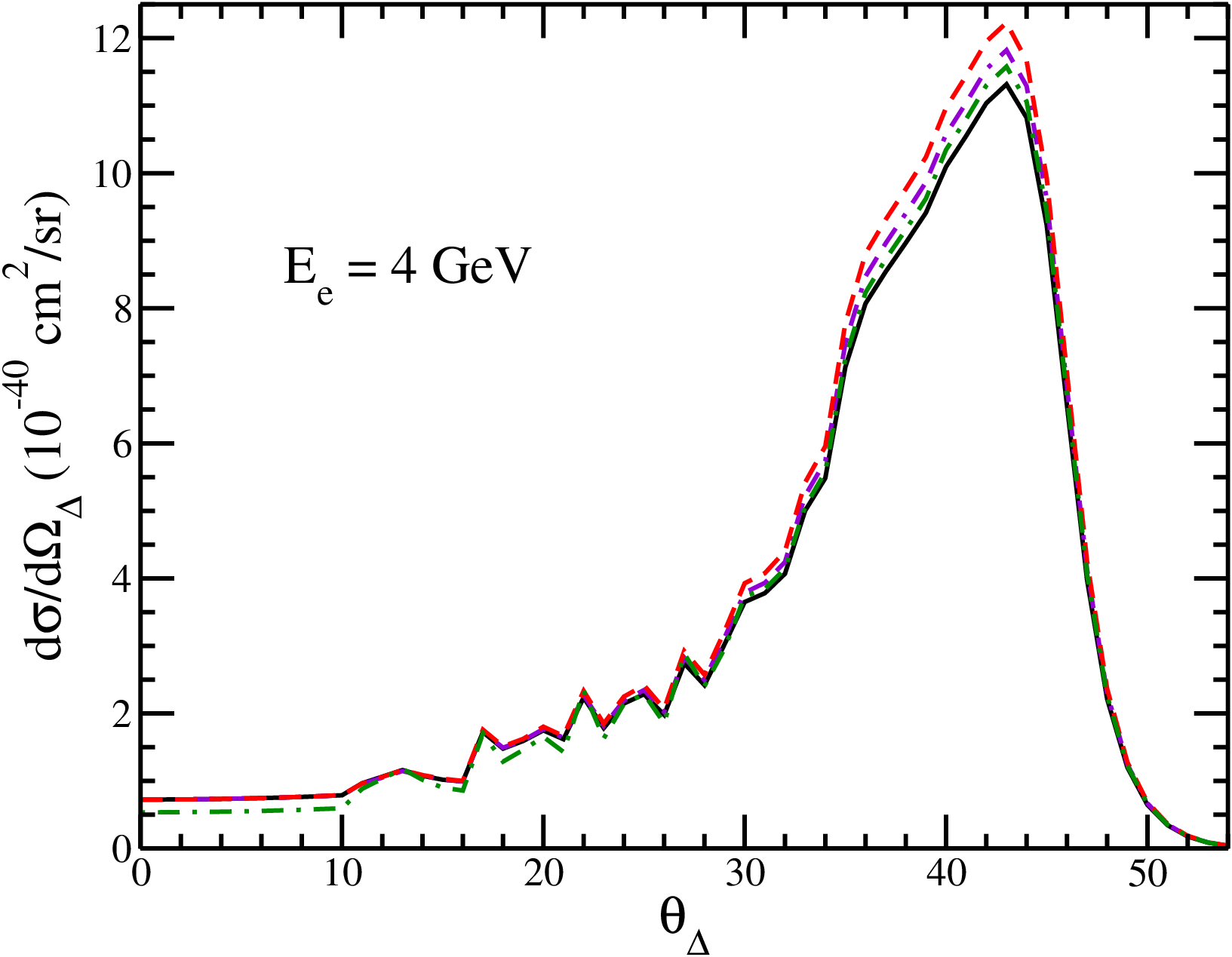}
\caption{$\frac{d\sigma}{d\Omega_{\Delta}}$ as a function of $\cos \theta_{\Delta}$ for the process $e^- + p \longrightarrow \nu_e + \Delta^0$ at different values of electron energies viz. $E_{e}=500$~MeV~(upper left panel), 1~GeV~(upper right panel), 2~GeV~(lower left panel), and 4~GeV~(lower right panel). Lines and points have the same meaning in Fig.~\ref{sigma:delta_weak}.}
\label{fg:dsigma_weak_electron}
\end{center}
\end{figure}

\begin{figure}
\begin{center}
\includegraphics[width=0.47\textwidth,height=.45\textwidth]{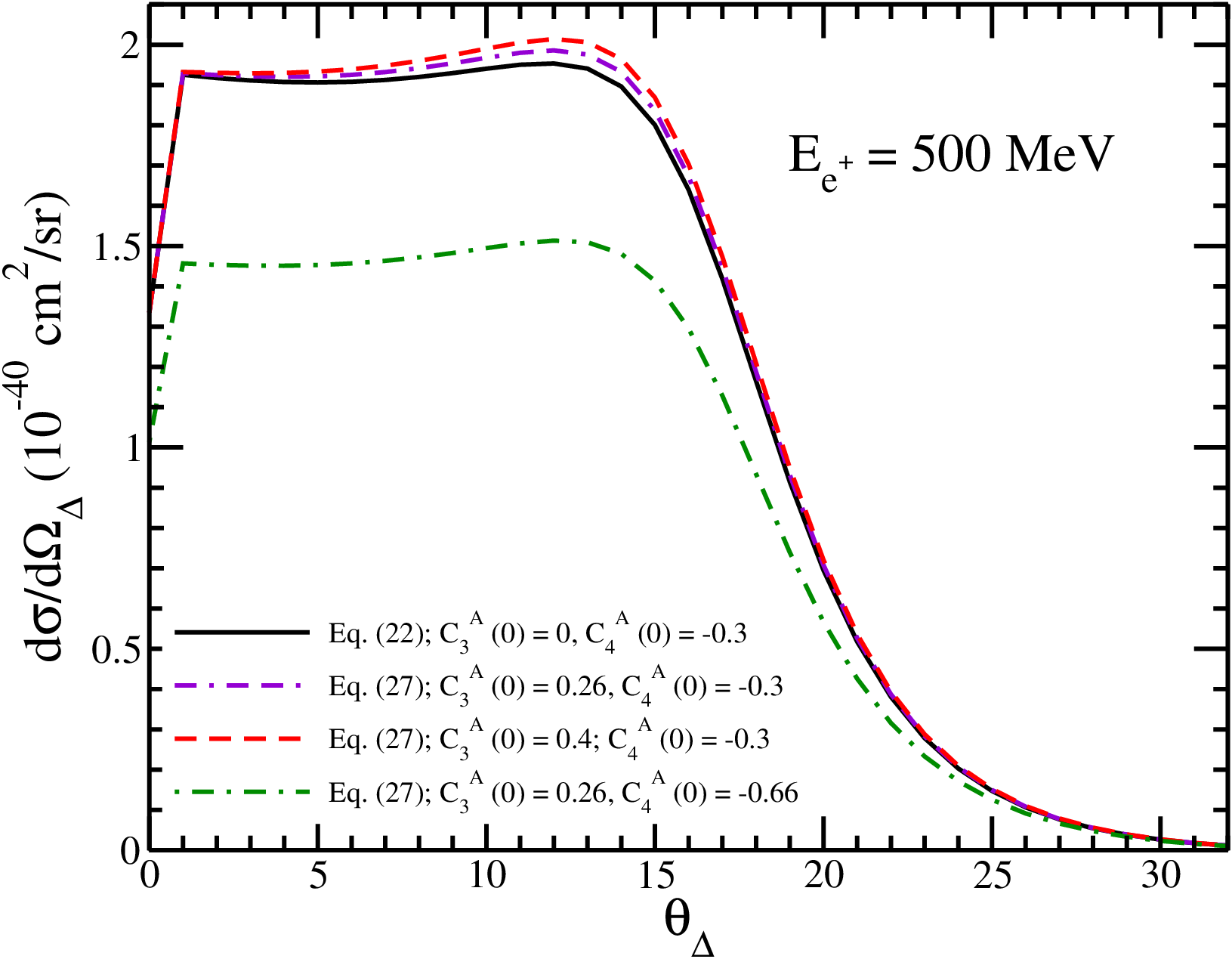}
\includegraphics[width=0.47\textwidth,height=.45\textwidth]{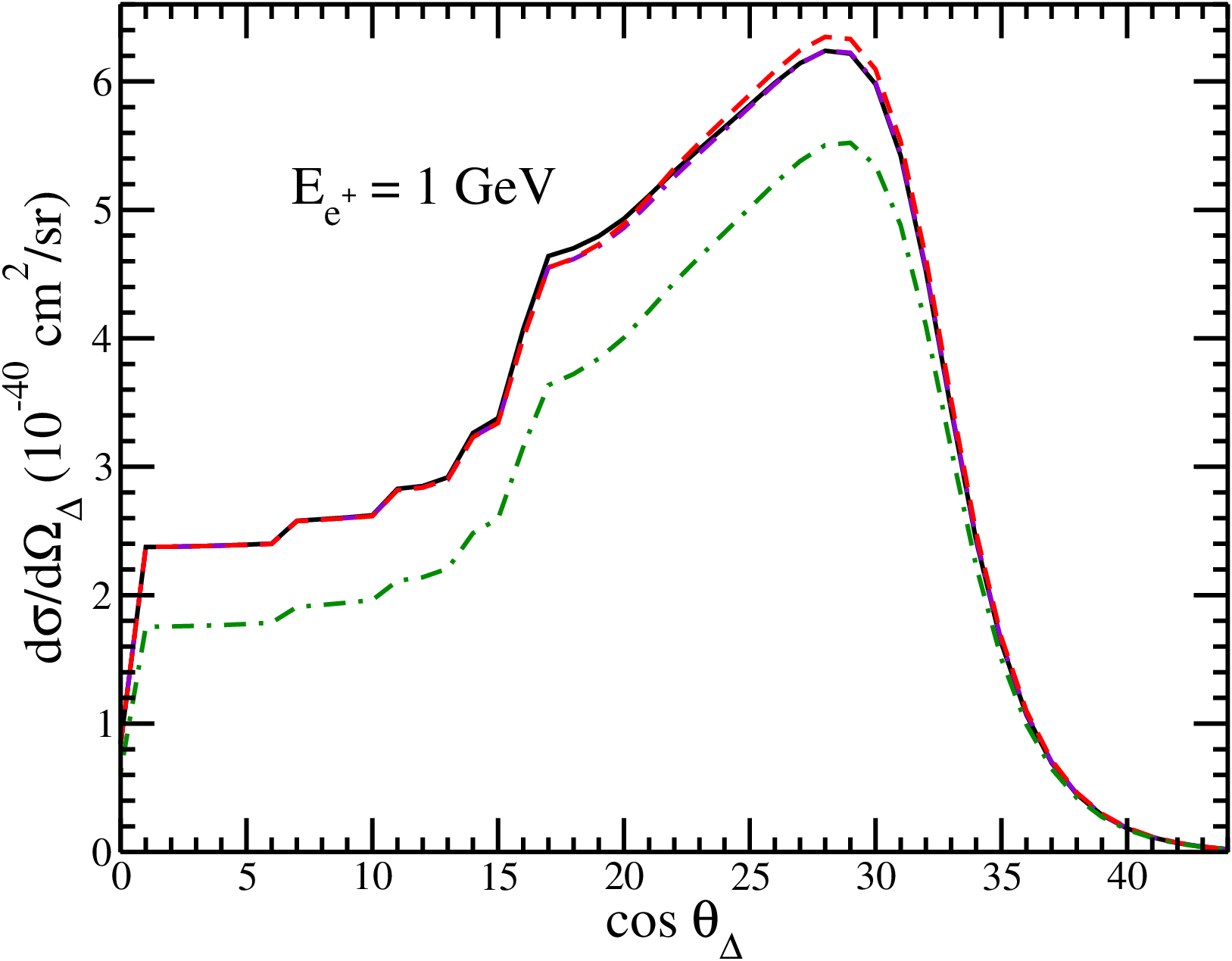}
\includegraphics[width=0.47\textwidth,height=.45\textwidth]{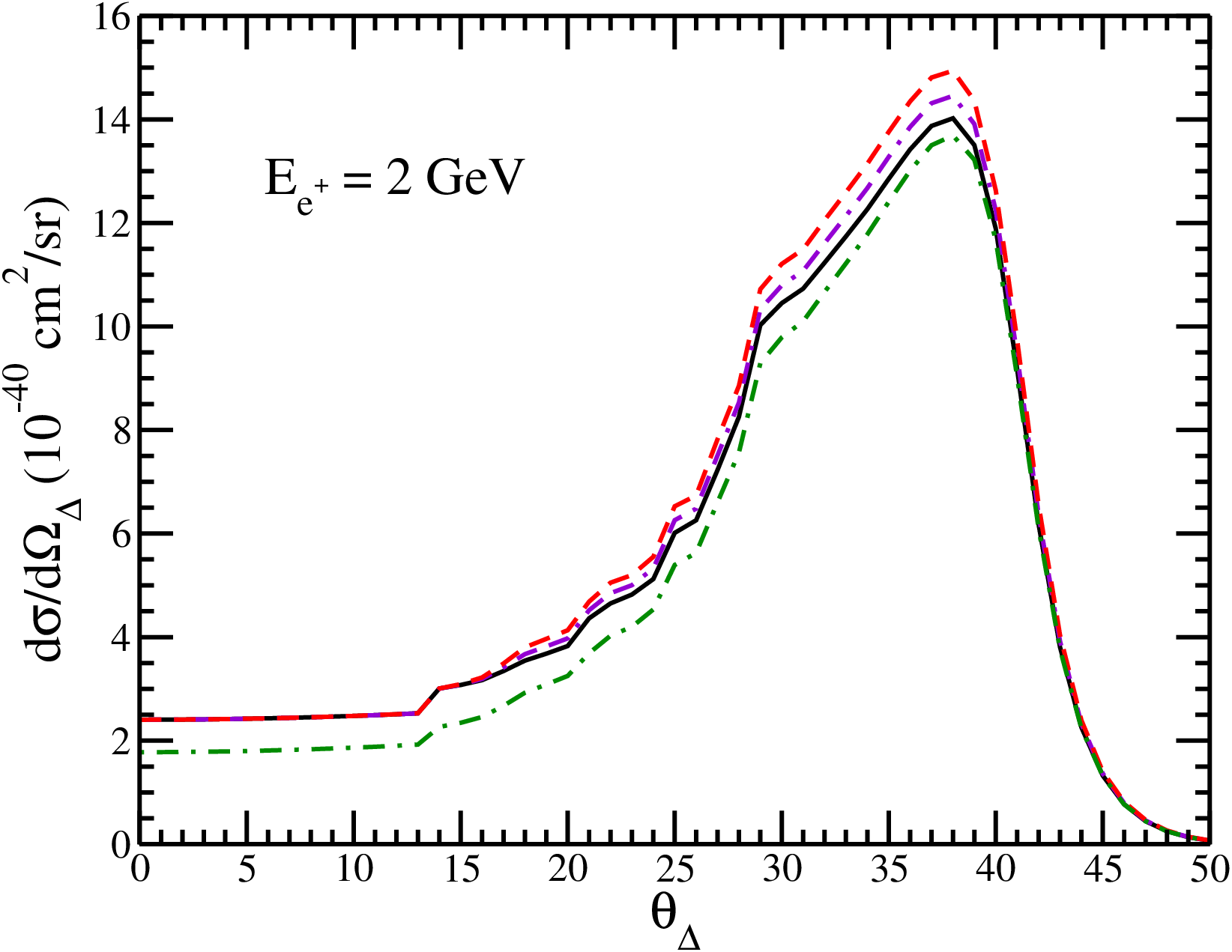}
\includegraphics[width=0.47\textwidth,height=.45\textwidth]{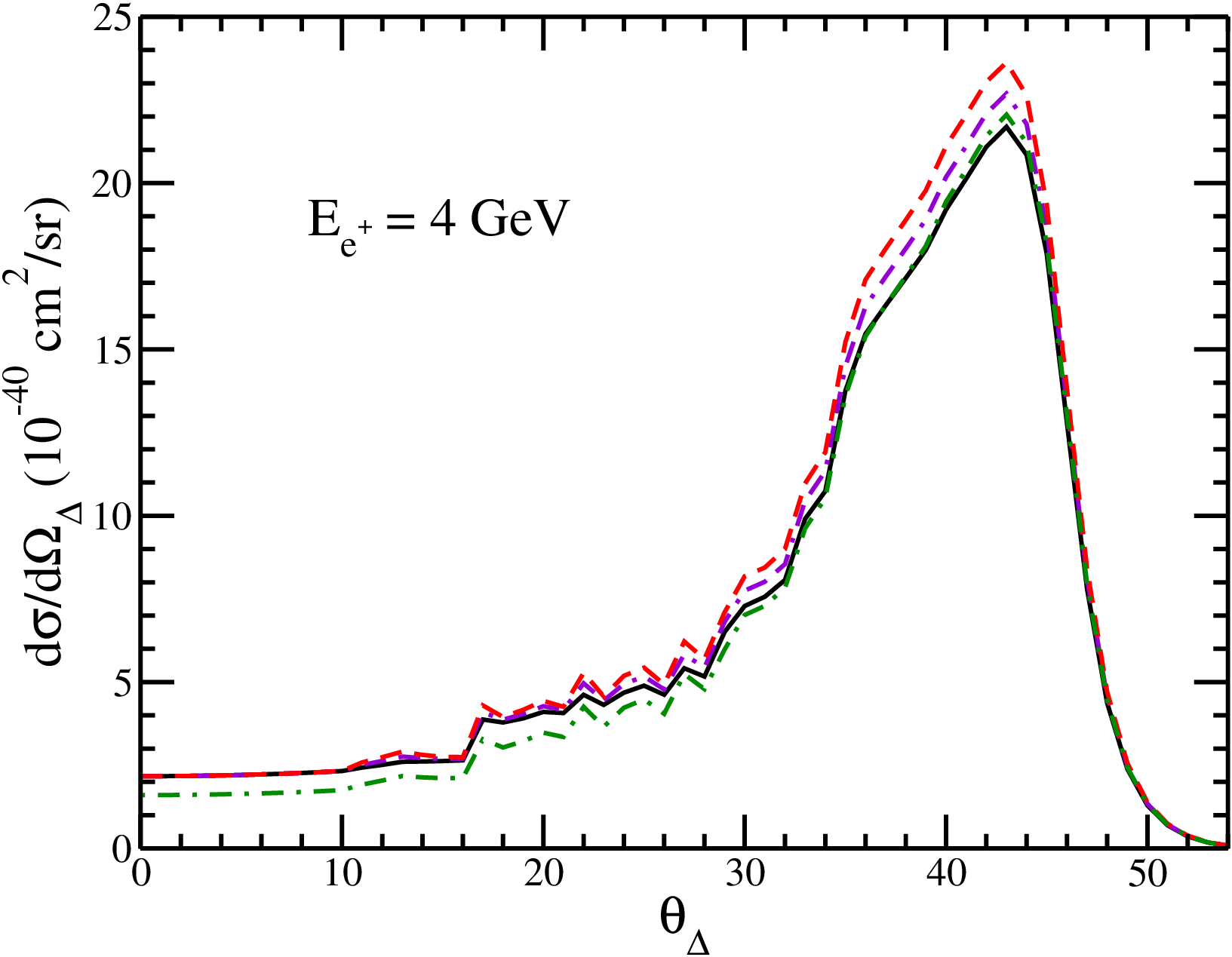}
\caption{$\frac{d\sigma}{d\Omega_{\Delta}}$ as a function of $\cos \theta_{\Delta}$ for the process $e^+ + p \longrightarrow \bar{\nu}_e + \Delta^{++}$ at different values of electron energies viz. $E_{e}=500$~MeV~(upper left panel), 1~GeV~(upper right panel), 2~GeV~(lower left panel), and 4~GeV~(lower right panel). Lines and points have the same meaning in Fig.~\ref{sigma:delta_weak}.}
\label{fg:dsigma_weak_positron}
\end{center}
\end{figure} 

In Fig.~\ref{fg:dsigma_nu_weak_electron_C3band}, we present the results for $\frac{d\sigma}{dQ^2}$ vs. $Q^2$ for the process $e^{-} p \longrightarrow \Delta^{0}\nu_{e}$ at $E_{e}=500$~MeV and 1~GeV, to study the effect of $C_{3}^{A}(0)$ on the $Q^2$ distribution by 
fixing the value of $C_{4}^{A}(0)=-0.66$ and using $C_{3}^{A}(0) =0.26^{+0.17}_{-0.04}$ given by Chen et al.~\cite{Chen:2023zhh}.
These results are compared with the results obtained using Adler's model i.e., $C_{3}^{A}(0)=0$, $C_{4}^{A}(0)=-0.3$. To quantify the effect of $C_{3}^{A}(0)$, we take the central value of Chen et al.~\cite{Chen:2023zhh}~($C_{3}^{A}(0) = 0.26$) keeping $C_{4}^{A}(0) = -0.3$.   We find that a non-zero value of $C_{3}^{A}(0)$ results in increasing the differential cross section, which is around 8--10\% in the peak region of $Q^2$. 
With this value of $C_{3}^{A}(0)$, we change $C_{4}^{A}(0)$~(shown by the dashed line) following Chen et al.~\cite{Chen:2023zhh} and find a reduction in the $\frac{d\sigma}{dQ^2}$, which is about 10--15\% in the peak region. 
We also show through the band in the figure, which represents the results obtained for the two extreme values of $C_{3}^{A}(0)$ in the Chen's model~\cite{Chen:2023zhh}, which leads to a variation of about $5\%$ in the $Q^2$ distribution. 

For the two different parameterizations of $C_{4}^{A} (Q^2)$ used in this work,  we present, in Fig.~\ref{fg:dsigma_nu_weak_electron_C4band}, the results for $\frac{d\sigma}{dQ^2}$ vs. $Q^2$ for the process $e^{-} p \longrightarrow \Delta^{0}\nu_{e}$ at $E_{e}=500$~MeV and 1~GeV, by taking the prescription of Chen et al.~\cite{Chen:2023zhh}~(top panel) and Graczyk et al.~\cite{Graczyk:2009zh}~(bottom panel), and compare them with the results obtained using the Adler's model~\cite{Adler:1968tw}.
To quantify the effect of $C_{4}^{A} (0)$, we present the result for $C_{3}^{A}(0)=0$ and $C_{4}^{A}(0)=-0.66$, corresponding to the central value of $C_{4}^{A}(0)$ in the Chen's model~\cite{Chen:2023zhh}, and compare them with the results obtained in  the Adler's model~\cite{Adler:1968tw}. 
We find a reduction in the $Q^{2}$ distribution, which is around 20$\%$ in the peak region of $Q^2$ and becomes smaller with the increase in $Q^2$. 
Then to show the variation in the $Q^2$ distribution for a finite $C_{3}^{A}(0)$ in the model of Chen et al.~\cite{Chen:2023zhh}, we have shown the results with the dashed line when $C_{3}^{A}(0)=0.26$ and $C_{4}^{A}(0)=-0.66$ and the band corresponds to the two extreme values of $C_{4}^{A} (0)$. 
We find that the cross section increases for a non-zero value of $C_{3}^{A}(0)=0.26$. With $C_{3}^{A}(0)=0.26$ and varying $C_{4}^{A}(0)$ between the two extreme values, we find a variation of $< 5\%$ in the peak region of $Q^2$, which becomes smaller for the lower and higher values of $Q^2$. 
Using the parameterization of Graczyk et al.~\cite{Graczyk:2009zh}~(bottom panel), with $C_{3}^{A}(0)=0$ and $C_{4}^{A} (0) = -0.67 \pm 0.42$, we find a broader band, the width of which decreases with the increase in electron energy. 
It must be pointed out that the results shown by the dashed-dotted line and by the solid line with diamonds are obtained at similar values of $C_{4}^{A}(0) = -0.66$, however, with two different parameterizations of $C_{4}^{A}(Q^2)$. 

Similar study has been made for a positron beam, and in Fig.~\ref{fg:dsigma_nu_weak_positron_C3band}, we present the results for $\frac{d\sigma}{dQ^2}$ vs $Q^2$ at $E_{e}=500$~MeV, 1~GeV, 2~GeV, and 4~GeV, by varying the value of $C_{3}^{A}(0)=0.26^{+0.17}_{-0.04}$, as calculated in the model of Chen et al.~\cite{Chen:2023zhh}~(Eq.~(\ref{c3A_Chen})) keeping $C_{4}^{A}(0)=-0.66$. 
These results are compared with the results obtained using the form factor parameterization given by Adler~\cite{Adler:1968tw}~(Eq.~(\ref{cia_lala})).
We find that a non-zero value of $C_{3}^{A}(0)=0.26$, with $C_{4}^{A}(0)=-0.3$, has almost no effect on the differential cross section at all values of the positron energies. 
However, changing the value of $C_{4}^{A} (0)$ from $-0.3$ to $-0.66$~(as predicted in the model of Chen et al.~\cite{Chen:2023zhh}) with fixed $C_{3}^{A}(0)=0.26$, we observe a reduction in  the differential cross section which is about 20$\%$ at $Q^2=0$, which  gradually decreases with the increase in $Q^2$ for all values of positron energies. Varying $C_{3}^{A}(0)$ in the given limits results a very small change in the $Q^2$ distribution, which is $1-2\%$ for $E_{e^{+}} \le 1$~GeV, and becomes $3-4\%$ for $E_{e^{+}} = 4$~GeV. 

To observe the effect of $C_{4}^{A}(Q^2)$ on the $Q^2$ distribution in Figs.~\ref{fg:dsigma_nu_weak_positron_C4band_Chen} and \ref{fg:dsigma_nu_weak_positron_C4band_Graczyk},  respectively, we present the results for $\frac{d\sigma}{dQ^2}$ vs $Q^2$ at $E_{e}=500$~MeV, 1~GeV, 2~GeV, and 4~GeV for the $e^+ + p \longrightarrow \bar{\nu}_e + \Delta^{++}$ scattering process, using the parameterization of $C_{4}^{A}(Q^2)$ given in the models of Chen et al.~\cite{Chen:2023zhh} and Graczyk et al.~\cite{Graczyk:2009zh}. 
It may be observed that~(Fig.~\ref{fg:dsigma_nu_weak_positron_C4band_Chen}), the value of $C_{4}^{A}(0)=-0.66^{+0.03}_{-0.10}$ predicted in the model of Chen et al.~\cite{Chen:2023zhh} leads to  a small change in the differential cross section for a finite $C_{3}^{A}(0)=0.26$, which is about 8$\%$ at $Q^2=0$, and becomes small with the increase in $Q^2$.
On the other hand, using the parameterization of $C_{4}^{A}(Q^2)$ given by Graczyk et al.~\cite{Graczyk:2009zh}~(Fig.~\ref{fg:dsigma_nu_weak_positron_C4band_Graczyk}), we find a net variation of about 48$\%$ in the peak region of $Q^2$ which becomes smaller with the increase in $Q^2$ and the difference diminishes for $Q^2 > 0.4$GeV$^2$. Then to show the variation in the $Q^2$ distribution for a finite $C_{3}^{A}(0)$, we have shown the results with the dashed line when $C_{3}^{A}(0)=0.26$ and $C_{4}^{A}(0)=-0.66$ and the band corresponding to the dashed line is for $C_{4}^{A}(0)=-0.69$ and $C_{4}^{A}(0)=-0.56$. Keeping $C_{3}^{A}(0)=0.26$, we find a change of about 10$\%$ at $Q^2=0$ in the differential cross section when $C_{4}^{A}(0)$ is varied by about 8$\%$ at $Q^2=0$, which becomes small with the increase in $Q^2$.

Therefore, one may obtain information about the axial vector form factor $C_{4}^{A}(Q^2)$ independent of $C_{5}^{A} (Q^2)$, as suggested in the Adler's model~\cite{Adler:1968tw}, by studying the differential cross section for the positron induced $\Delta^{++}$ production as the contribution from $C_{3}^{A} (Q^2)$ is almost negligible in this case. Moreover, the differential cross section for the electron induced $\Delta^{0}$ production is sensitive to both $C_{3}^{A} (Q^2)$ and $C_{4}^{A} (Q^2)$ form factors.

\subsection{Experimental prospects of the weak production of $\Delta(1232)$}
In view of the availability of the electron/positron beams with the luminosities in the range of $10^{38}-10^{39}$~cm$^{-2}$~sec$^{-1}$, it should be feasible to observe the weak production of $\Delta(1232)$ induced by the electrons and positrons. 
We, therefore, discuss the prospects of observing these reactions at JLab and the possibility of determining the $N-\Delta$ transition axial vector form factors $C_{i}^{A}(Q^2);~(i=3-5)$.
For this, we purpose, we have shown in Figs.~\ref{fg:dsigma_weak_electron} and \ref{fg:dsigma_weak_positron}, the numerical results for the angular distribution $\frac{d\sigma}{d\Omega_{\Delta}}$ vs. $\theta_{\Delta}$ for the various energies of the electrons and positrons in the range 0.5--4~GeV. 
We see, from these figures, that
\begin{itemize}
 \item [(i)] The differential cross section $\frac{d\sigma}{d\Omega_{\Delta}}$ lies in the range of $10^{-40}-10^{-39}$~cm$^{2}/$sr for the electron and positron induced $\Delta$ production in the peak region.
 
 \item [(ii)] The differential cross section $\frac{d\sigma}{d\Omega_{\Delta}}$ for the positron induced production is generally larger~(smaller) than the electron induced production cross sections for energies larger~(smaller) than 1.5~GeV. The positron induced production cross section becomes two times large in the peak region around $E_{e}=4$~GeV.
 
 \item [(iii)] The differential cross section $\frac{d\sigma}{d\Omega_{\Delta}}$ for both processes peaks around $\theta_{\Delta}=25^{\circ}$ at $E_{e} = 1$~GeV, which shifts to higher angles i.e. $35^{\circ}$ and $40^{\circ}$ for $E_{e}=2$ and 4~GeV, respectively.
 
 \item [(iv)] The differential cross section $\frac{d\sigma}{d\Omega_{\Delta}}$ for both processes remains of the order of $10^{-40}-10^{-39}$~cm$^{2}/$sr in a wide range of $\theta_{\Delta}$, which is around $15^{\circ}-20^{\circ}$ for the electron/positron energies in the region of $1-4$~GeV.
\end{itemize}

With these characteristics of $\frac{d\sigma}{d\Omega_{\Delta}}$ vs. $\theta_{\Delta}$ for the electron/positron induced $\Delta(1232)$ production processes, we estimate the count rates for a detector subtending a solid angle $\Delta\Omega_{\Delta}$ as:
\begin{equation}
 \text{counts/hr} = \frac{d\sigma}{d\Omega_{\Delta}} \Delta\Omega_{\Delta} \times \text{Luminosity}~(L) \times \text{ detector efficiency } \times 3600.
\end{equation}
Using $\Delta\Omega_{\Delta} = 2\pi \sin\theta_{\Delta} \cdot \Delta \theta_{\Delta}$~(in degrees)~$\times \frac{\pi}{180}$, luminosity $L \approx 10^{39}$~cm$^{-2}$~sec$^{-1}$, and $\frac{d\sigma}{d\Omega_{\Delta}} \approx 10^{-39}$~cm$^{2}$, we obtain
\begin{equation}
 \text{counts/hr} = 400 \times \sin \theta_{\Delta} \times \text{detector efficiency/hr},
\end{equation}
where $\theta_{\Delta}$ is in degrees.

This estimate of counts/hr for the weak production of $\Delta$ is high enough to motivate the experiments to be undertaken to observe the weak production of $\Delta$ at JLab with continuous beams, which can be available for longer duration of time. 
It has been reported that the G0 experiment for studying the parity violation in the inelastic production of $\Delta$ has a running time of around 700 hrs~\cite{Souder:2015mlu}. 
With such exposure of electron/positron beams and increased target thickness, the above count rate can be further enhanced.

In the experimental study of the electromagnetic and weak neutral current production of $\Delta$ induced by the electrons $ep \longrightarrow e\Delta^{+}$  by Wang et al.~\cite{Wang:2013kkc, Wang:2014guo, PVDIS:2014cmd} and Androic et al.~\cite{Qweak:2013zxf, QWeak:2019kdt}, respectively, the reconstruction of $\Delta$ have been done quite successfully by the observation of the nucleons and pions produced in the final state. 
Therefore, it should also be possible to do this identification in the case of the charged current induced production of $\Delta(1232)$.
In the case of weak charged current production of $\Delta$, the final hadrons i.e., nucleon and pion events produced through the excitation and subsequent strong decays of $\Delta$ are distinct from the electromagnetic and weak neutral current events as they do not have a charged lepton in the final state. 
However, these reactions produce $\Delta^{+}$ while the charged current reactions produce $\Delta^{0}$ leading to different charged states of nucleons and pions and  therefore, the weak production of $\Delta(1232)$ is quite distinct from the electromagnetic production of $\Delta(1232)$. 
Moreover, there could be background, in the case of electron induced $\Delta(1232)$ production, due to the nucleons and pions produced through the excitation of $\Lambda^{0}$ and $\Sigma^{0}$ through the weak processes of $e^{-} + p \longrightarrow \Lambda^{0}(\Sigma^{0}) + \nu_{e}$ and their subsequent decays, as the production cross section of $\Lambda^{0}(\Sigma^{0})$ is also of the same order as that of the $\Delta$ production~\cite{Akbar:2017qsf}. 
However, the nucleons and the pions produced by the decays of $\Lambda^{0}(\Sigma^{0})$ are well separated from those produced by the $\Delta$ decays. 
This is because the $\Delta$ decays via the strong interaction and occur immediately at the interaction point while $\Lambda^{0}$ and $\Sigma^{0}$ decays being weak decays occur at a later time allowing the $\Lambda^{0}(\Sigma^{0})$ to travel some distance. 
Thus, reconstruction of the nucleons and pions from the vicinity of the interaction point would lead to $\Delta (1232)$ as in the case of electromagnetic and weak neutral current production of $\Delta$. 
The energy resolution required to identify $\Delta(1232)$ is already available at the JLab and Mainz laboratories in the respective energy ranges of $\Delta$ production.
There are no such background problems in the case of positron induced $\Delta (1232) $ production as they lead to $\Delta^{++}$ state, which decays into a proton and a $\pi^{+}$, i.e., the two charged particles in the final state. 

In view of the above discussion, it seems quite feasible to experimentally observe the weak production of $\Delta$ induced by the electrons and positrons. 
Therefore, it seems possible to determine the axial vector form factors $C_{i}^{A} (Q^2); (i=3-5)$ including the subdominant form factors $C_{3}^{A} (Q^2)$ and $C_{4}^{A} (Q^2)$ by analyzing $\frac{d\sigma}{d\Omega_{\Delta}}$ vs. $\theta_{\Delta}$ in the peak region corresponding to higher $\theta_{\Delta}$~(see Figs.~\ref{fg:dsigma_weak_electron} and \ref{fg:dsigma_weak_positron}).

\section{Summary and Conclusions}\label{summary}
This work presents a study of the charged current weak production of $\Delta(1232)$ induced by the electrons and positrons from the proton target. 
The numerical results for the total scattering cross section~$\sigma (E)$, differential scattering cross section~$\frac{d\sigma}{dQ^2}$, and the angular distribution $\frac{d\sigma}{d\Omega_{\Delta}}$ for the various energies of electrons and positrons in the range 0.5--4~GeV are discussed by considering all the $N-\Delta$ transition vector $C_{i}^{V} (Q^2); (i=3-5)$ and axial vector $C_{i}^{A} (Q^2); (i=3-5)$ form factors in the weak $N-\Delta$ transition amplitude. 
For the isovector vector $C_{i}^{V} (Q^2); (i=3-5)$ and the axial vector $C_{5}^{A} (Q^2)$ form factors, we use the parameterization of Lalakulich et al.~\cite{Lalakulich:2005cs}.
We have considered the different parameterizations for the subdominant axial vector form factor $C_{3}^{A} (Q^2)$ and $C_{4}^{A} (Q^2)$ given by the phenomenological studies by Adler~\cite{Adler:1968tw} and Graczyk et al.~\cite{Graczyk:2009qm, Graczyk:2009zh} as well as by the theoretical studies of Chen et al.~\cite{Chen:2023zhh} and Barquilla-Cano et al.~\cite{Barquilla-Cano:2007vds} parameterized by us.

To summarize our results, we find:
\begin{itemize}
\item [(i)] The cross sections $\sigma(E)$, and the differential cross sections $\frac{d\sigma}{dQ^2}$ in the peak region of $Q^2$ are found to be of the order of $10^{-39}-10^{-40}$~cm$^{2}$. 
With the availability of continuous electron/positron beams with a luminosity of the order of $10^{38}-10^{39}$~cm$^{-2}$~sec$^{-1}$ available at JLab and Mainz, it should be possible to observe the weak charged current production of $\Delta$ by doing long exposure experiments for determining the weak axial vector form factors $C_{i}^{A} (Q^2); (i=3-5)$. 

 \item [(ii)] The production cross section for the electron induced process increases with the increase in incoming electron energy and saturates at around 2~GeV while in the case of positron induced process, the production cross section increases with the increase in positron energy, even at $E_{e^{+}}=4$~GeV and saturates at around 8~GeV.
 
 \item [(iii)] In the region of $E_{e}<1.5$~GeV, the production cross section for the electron induced process is more as compared to the positron induced process, while for $E_{e}>1.5$~GeV, the production cross section for the positron induced process becomes larger. 
 
 \item [(iv)] In the case of electron induced $\Delta^{0}$ production process, the non-zero value of $C_{3}^{A} (0)$ increases the differential scattering cross section $\frac{d\sigma}{dQ^2}$, while a value of $C_{4}^{A} (0)$ smaller than the value suggested in the Adler's model but predicted in the models of Graczyk et al. and Chen et al., decreases $\frac{d\sigma}{dQ^2}$. 
 The combination of these effects, for example, in the model of Chen et al., reduces the differential cross section as compared to the one calculated in the Adler's model. 
  
 \item [(v)] In the case of positron induced $\Delta^{++}$ production cross section, we find almost no effect of a non-zero value of $C_{3}^{A} (0)$ on the differential cross section, especially in the low energy region of positron i.e. $E_{e^{+}} < 1$~GeV. 
 However, using a smaller value of $C_{4}^{A}(0)$, calculated in the model of Chen et al., as compared to the one suggested by the Adler's model, we observe a significant decrease in $\frac{d\sigma}{dQ^2}$ in the peak region of $Q^2$ at all values of the positron energies considered in this work.
 
 \item [(vi)] In the case of angular distribution of $\Delta(1232)$ i.e., $\frac{d\sigma}{d\Omega_{\Delta}}$, the cross sections in the peak region of $\theta_{\Delta}$ is found to be of the order of $10^{-40}-10^{-39}$~cm$^{2}$, and remains sufficiently large in a wide $\theta_{\Delta}$ range of about $15^{\circ} - 20^{\circ}$, making it accessible to the experimental observation.  
\end{itemize}

We conclude that, with the availability of the intense electron and positron beams with very high luminosities at JLab, it may be possible to observe the weak charged current production of $\Delta(1232)$ induced by the electrons and positrons, and determine the axial vector $N-\Delta$ transition form factors $C_{i}^{A} (Q^2); (i=3-5)$. 
The cross sections obtained with the electron/positron beams with well defined energy and direction will be free from any uncertainties arising due to the incident beam flux as in the case of neutrino and antineutrino experiments. 

 \section*{Acknowledgements}
AF and MSA are thankful to the
Department of Science and Technology (DST), Government of India for providing financial assistance under Grant No.
SR/MF/PS-01/2016-AMU.

\end{document}